\documentclass[12pt,aps,preprint,titlepage,nofootinbib,noshowkeys,noshowpacs,longbibliography,superscriptaddress,tightenlines]{revtex4-2}
\usepackage{amsmath,amsfonts,amsthm,amssymb}
\usepackage{graphics,graphicx}
\usepackage[inline]{enumitem}
\usepackage{mathrsfs}
\usepackage{bm}
\usepackage{color}

\definecolor{darkblue}{RGB}{0,0,196}
\definecolor{darkgreen}{RGB}{0,120,0}

\usepackage{stackengine}

\def\bb{{(b)}}

\def\edense{{\mathcal{E}}}
\def\pdense{M}
\def\dpdense{\delta\kern-0.08em M}
\def\dbeta{\delta\kern-0.08em \beta}

\newcommand{\Kinf}{\mathcal{K}}   %
\newcommand{\Ce}{C_e}   %
\newcommand{\cRTA}{c}             %
\newcommand{\aone}{a_1}
\newcommand{\atwo}{a_2}
\def\Cinf{C_{\infty}}

\def\noisekernel{{\kappa}}
\def\Eq#1{eq.~(\ref{#1})}

\def\eq#1{(\ref{#1})}
\def\app#1{App.~\ref{#1}}
\def\Fig#1{Fig.~\ref{#1}}

\def\Sect#1{Sect.~\ref{#1}}

\def\Kd{{K^\tau}}
\def\beq{\begin{equation}}
\def\eeq{\end{equation}}
\def\st{\begin{equation}}
\def\stp{\end{equation}}
\def\ba{\begin{eqnarray}}
\def\ea{\end{eqnarray}}

\def\dd{{\rm d}}

\def\tideal{{\mathcal T}}
\def\tvisc{\Pi}

\def\bb1{{(1)}}

\usepackage[colorlinks=true,linktocpage=true,linkcolor=darkblue,citecolor=red,urlcolor=darkblue]{hyperref}
\usepackage{cleveref}

\begin{document}
\preprint{}
 
\title{Relativistic viscous hydrodynamics in the Density Frame for heavy-ion collisions}

\author{Jean-Fran\c{c}ois Paquet}
\email{jean-francois.paquet@vanderbilt.edu}
\affiliation{Department of Physics and Astronomy, Vanderbilt University, Nashville TN 37240}

\author{Mayank Singh}
\email{mayank.singh@vanderbilt.edu}
\affiliation{Department of Physics and Astronomy, Vanderbilt University, Nashville TN 37240}

\author{Derek Teaney}
\email{derek.teaney@stonybrook.edu}
\affiliation{Center for Nuclear Theory, Department of Physics and Astronomy, Stony Brook University, Stony Brook, New York, 11794-3800, USA}

\date{\today}

\begin{abstract}
We present a numerical study of relativistic viscous hydrodynamics in the Density Frame for
boost-invariant flows in 2+1 dimensions. The Density Frame is a formulation of relativistic
viscous hydrodynamics that is first order in time, requires no auxiliary fields, and has no
non-hydrodynamic modes; the only parameters are the shear and bulk viscosities and the
equation of state. We formulate the numerical method in terms of a discrete analog of the
dissipative function, which guarantees that the discrete equations produce entropy on all
length scales and leads to a remarkably robust algorithm. We test the method against the exact solution of kinetic
theory in the relaxation time approximation with Gubser symmetry. In making this comparison,
we clarify how the kinetic solution matches onto hydrodynamics: we organize the solution by
the opacity and show how the Bjorken attractor passes onto the Gubser solution order by order
in gradients. We then replace the hydrodynamic engine of the MUSIC code with the Density
Frame, and compare the two for IP-Glasma initial conditions in Pb-Pb collisions at
$\sqrt{s_{NN}} = 5.02$~TeV. The Density Frame smooths sharp gradients more rapidly. The
largest differences arise in the bulk-viscous sector, where the bulk pressure in the Density
Frame responds instantaneously to the flow, while in the M\"uller-Israel-Stewart-based MUSIC it relaxes over a finite time. With
shear viscosity alone, the multiplicity, mean transverse momentum, and elliptic flow of final-state hadrons from the
two codes are very similar. The Density Frame is now ready for a complete Bayesian analysis of
heavy-ion data.
\end{abstract}

\maketitle

\section{Introduction} 
\label{sec:intro}

\subsection{Hydrodynamics, transport coefficients, and the Density Frame}

Nuclear collisions at the Relativistic Heavy Ion Collider (RHIC) and the Large Hadron
Collider (LHC) exhibit remarkable patterns of collective flow, which are described by
relativistic viscous hydrodynamics~\cite{Heinz:2013th,Gale:2013da,Teaney:2009qa,Busza:2018rrf,Heinz2026}.
Through multistage simulations of heavy-ion collisions built around hydrodynamics, these flow
patterns constrain the transport properties of the quark-gluon plasma and its equation of state.
Bayesian
comparisons between simulation and experiment now give increasingly quantitative estimates of
the shear and bulk viscosities of QCD~\cite{Bernhard:2019bmu,PhysRevLett.126.202301,JETSCAPE,Heffernan:2023utr,Paquet:2023rfd},
and flow-like correlations are seen even in proton-nucleus and other small
systems~\cite{CMS:2015yux,PHENIX:2015idk}.

There are, however, conceptual challenges with the hydrodynamic theories used in these
simulations. Relativistic Navier--Stokes theory in the Landau or Eckart frame is unstable and
acausal~\cite{Hiscock:1985zz}. The standard remedy, due to M\"uller and to Israel and
Stewart (MIS)~\cite{Muller1967,Israel:1976tn,Israel:1979wp}, promotes the viscous stresses to
dynamical fields which relax to their Navier--Stokes values over a time $\tau_\pi$.
Nearly all simulations of heavy-ion collisions use a variant of this approach, often
called second-order hydrodynamics~\cite{Baier:2007ix,Denicol:2012cn}. The more recent
formulation of Bemfica, Disconzi, Noronha and Kovtun (BDNK)~\cite{Bemfica:2017wps,Kovtun:2019hdm}
adds no new fields, but its equations are second order in time and depend on additional
frame parameters. In each case the extra freedom introduces non-hydrodynamic modes. These modes are needed to preserve causality, but they model the microscopic physics in a
non-universal way and lie outside the realm of hydrodynamics.
The associated model parameters, such
as $\tau_\pi$ in MIS theory or the frame parameters of BDNK, can obscure which universal transport
properties are actually being extracted from data, especially in small systems.

The Density Frame formulation of viscous hydrodynamics has no additional non-hydrodynamic
modes or parameters, resolving these
conceptual and practical issues~\cite{Armas:2020mpr,Basar:2024qxd,TheoryPaper,Bhambure:2024axa}. 
Given a foliation of
spacetime, the conserved energy and momentum densities on each spatial slice,
$\edense = T^{00}$ and $\pdense^i = T^{0i}$, are the slow variables of the effective theory.
They are evolved with the conservation laws, and all other quantities, including the spatial
stress $T^{ij}$, are determined instantaneously by $\edense$ and $\pdense^i$ through the
gradient expansion. The theory is strictly first order in gradients, and the only parameters
are the shear and bulk viscosities and the equation of state. The equations are not manifestly
covariant, but they are covariant to a given order in the gradient expansion, provided a Lorentz
transformation is accompanied by a change of hydrodynamic frame~\cite{Armas:2020mpr,Basar:2024qxd}.

Although it is not the focus of this work, the Density Frame has an added benefit: the
equations fit neatly into the traditional framework of dissipative stochastic
processes~\cite{LandauStatPhys1,landau2013fluid,FoxUhlenbeck,FoxUhlenbeck2}. Noise can be added with
Metropolis steps~\cite{Basar:2024qxd,TheoryPaper}, an approach that has proved useful for
simulating systems close to critical points~\cite{Florio:2021jlx,Florio:2023kmy,Chattopadhyay:2023jfm,Chattopadhyay:2024jlh,Chattopadhyay:2024bcv}
and ordinary stochastic fluids~\cite{Harhoff:2026oxb}. We have previously studied the Density Frame
numerically for the stochastic relativistic advection-diffusion equation~\cite{Basar:2024qxd}.

\subsection{This work}

The deterministic version of Density Frame Hydrodynamics was studied previously in 1+1 dimensions, where we compared it with
QCD kinetic theory, BDNK, and MIS (MUSIC)~\cite{Bhambure:2024axa}.
In this paper, we extend these numerical methods and we solve the Density Frame equations in 2+1 dimensions in Bjorken coordinates,
establishing the Density Frame as a fully developed tool for studying heavy-ion collisions.
In formulating the
method we use a traditional concept, the dissipative
function~\cite{LandauStatPhys1,landau2013fluid,DZYALOSHINSKII198067}, which is the positive
semi-definite quadratic form in the strains that determines the rate of entropy production. We find that it has a discrete analog: the viscous stresses are derivatives of a
discrete dissipative function, so the semi-discrete viscous step produces entropy in the same way
as the continuum equations. This makes the method remarkably robust: we needed only a very small energy-density pedestal
to handle the dilute edge of the plasma, a level of regulation that we generally found
insufficient for MIS hydrodynamic solvers. The same methods
can be used in other domains, for example for viscous hydrodynamics in general
relativity~\cite{TheoryPaper}.

In this paper, we test the numerical scheme against the exact solution of RTA kinetic theory with Gubser
symmetry~\cite{Denicol:2014xca,Denicol:2014tha,Dash:2020zqx}. In making this comparison we
clarify how the kinetic solution matches onto hydrodynamics, which has not been fully
explained in the literature. We connect the Gubser kinetic solution to the opacity expansion
introduced in recent kinetic simulations~\cite{Kurkela2019flow,Ambrus2022range,Ambrus2022opacity},
and show how the Bjorken attractor~\cite{Heller2015attractor,Giacalone2019attractor} passes
smoothly onto the Gubser solution, order by order in gradients. We then use this
understanding to compare the kinetic solution with our numerical Density Frame solution.
The comparison also tests the approximate foliation independence (and by extension Lorentz invariance) of the Density Frame: in
Gubser's coordinates, with $\rho$ as the time variable, the fluid is at rest and the
Density Frame solution is Gubser's hydrodynamic solution. Our code, however, uses Bjorken $\tau$ as the
time variable, and agrees with Gubser's solution up to controlled gradient corrections. 

Finally, we have constructed a practical tool---packaged, public, and documented---so that it can replace the
hydrodynamic engine of existing heavy-ion codes~\cite{density-frame-github}. As an example, we
replaced the hydrodynamic engine of MUSIC~\cite{Schenke:2010nt,Schenke:2010rr,Paquet:2015lta}
with the Density Frame and compared the two for IP-Glasma initial
conditions~\cite{Schenke:2012wb}. 
The Density Frame rapidly smooths out very sharp gradients; this dissipation of fine structure
seems to us more faithful to the physics of viscous hydrodynamics. The largest difference is in the bulk-viscous sector:
in MIS simulations such as MUSIC the bulk pressure relaxes slowly near the crossover, while in the Density Frame it
responds instantaneously.
Building on these results, we show that simulations of heavy-ion collisions that replace MIS hydrodynamics with Density Frame lead to similar flow patterns and final hadronic observables.
In the future, this framework can be used to investigate the sensitivity to non-hydrodynamic modes in
small systems such as p+Pb and O+O collisions~\cite{Brewer:2021kiv}, and to address the theoretical
error bars in current Bayesian extractions of transport coefficients.

The paper is organized as follows. \Sect{sec:densityframe} presents the Density Frame
equations in Bjorken coordinates, the dissipative function, and the numerical method.
\Sect{sec:kinetic} compares the method with the Gubser kinetic solution, clarifying the matching between
the Bjorken attractor and three-dimensional hydrodynamics in the opacity expansion. \Sect{sec:music} compares the Density Frame with
MUSIC for realistic initial conditions, and \Sect{sec:conclusions} concludes.

\section{The Density Frame for 2+1D Bjorken Flow}
\label{sec:densityframe}

In this section we first review the Density Frame formulation of relativistic viscous
hydrodynamics, which is described in detail in our previous work~\cite{TheoryPaper,Bhambure:2024axa},
and then present the numerical method. We focus on the extension to Bjorken coordinates, and
initially describe the method in 1+1D for simplicity. With the right notation and concepts, the
extension to 2+1D is straightforward.
Relative to our previous work, the main new ingredient is to make maximal use of the dissipative function $R$~\cite{LandauStatPhys1,landau2013fluid,DZYALOSHINSKII198067}, which is the positive semi-definite quadratic form that appears in the entropy production.  We will see that 
it is possible to write down the dissipative function in a discrete form, and that the viscous strains can be implemented by taking derivatives of this function at the discrete level.  This approach is very effective and leads to a delightfully stable numerical method.

\subsection{The equations of motion}
\label{sec:eom}

Let us consider a special case of Bjorken flow with a non-zero flow velocity and spatial dependence only 
in the $x$ direction.
The metric in Bjorken coordinates is 
\st
   ds^2 = - \dd\tau^2 + (\dd x^2 + \dd y^2 + \tau^2 \dd\eta^2) \, ,
\stp
and the non-zero Christoffel symbols are $\Gamma^\tau_{\eta\eta} = \tau$ and
$\Gamma^\eta_{\tau\eta} = 1/\tau$.  

The normal vector for the foliation is simply $n^{\mu}=(1, 0, 0, 0)$
and the spatial metric is $h_{ij} = {\rm diag}(1, 1, \tau^2)$ where roman indices $i,j,k$ run over the spatial coordinates $x, y, \eta$. 
The volume element
of the spatial metric is $\sqrt{h} = \tau$. 
In two dimensions,  we will use $a,b,c$ for the transverse directions  $x,y$.
The foliation provided by the Bjorken
coordinates has extrinsic curvature $K_{ij} = -\nabla_i n_j = n_{\mu}\Gamma^{\mu}_{ij}$,
which has only one non-zero component $K_{\eta\eta} = -\Gamma^{\tau}_{\eta\eta}$.  To highlight the role of the extrinsic curvature we will write $\Gamma^{\tau}_{\eta\eta}$ instead of simply $\tau$ in the equations of motion.

The evolution variables in hydrodynamics are the energy and momentum densities 
\st
( T^{\tau\tau},  T^{\tau x} ) \equiv (\edense , \pdense^x)\,,
\stp
and the equations of energy-momentum conservation read
  \begin{subequations}
  \begin{align}
  \partial_\tau (\tau \edense) + \partial_x  (\tau \pdense^x)  + \tau \Gamma^{\tau}_{\eta\eta} T^{\eta\eta}=&  0 \, , 
   \label{eq:equation}   \\
  \partial_\tau  (\tau \pdense^x) + \partial_x (\tau T^{xx}) =&0 \,.  
  \label{eq:pequation}
\end{align} 
  \end{subequations}
These equations are extended to two dimensions in an obvious way by including the $y$ direction and the corresponding momentum density $\pdense^y$. The explicit factors of $\tau$ represent the volume element, $\sqrt{h}$.

The conservation equations are closed by specifying a
constitutive relation for the spatial stress tensor $T^{ij}$ in terms of
$(\edense, \pdense^x)$ and their spatial derivatives.
In ideal hydrodynamics this is  conveniently done by introducing intermediate parameters $\beta_{\mu}\equiv \beta u_{\mu}$ (the inverse temperature and flow velocity), which are defined by algebraic relations to the  conserved charge densities on the spatial slice:
\begin{subequations}
  \label{eq:densityframedef}
\begin{align}
  \edense =& \tideal^{\tau\tau}(\beta) \, , \\
  \pdense^x =& \tideal^{\tau x}(\beta)  \,  . 
\end{align}
\end{subequations}
The ideal stress tensor written here takes  
the functional form
\st
\tideal^{\mu\nu}(\beta) \equiv \left(e(\beta) + p(\beta)\right)  u^{\mu} u^{\nu} + p(\beta) g^{\mu\nu} \, . 
\stp
This equation  is to be understood as follows: $\beta_{\mu}$ is determined from the energy and momentum densities as specified by \Eq{eq:densityframedef},  and 
subsequently $\beta_{\mu}$ is used to specify the spatial stress 
\begin{align}
   \tideal^{ij}(\beta) =& \left(e(\beta) + p(\beta)\right)  u^{i} u^{j} + p(\beta) g^{ij} \, .
\end{align}
It is important that $T^{ij}$ is ultimately a function of $(\edense, \pdense^x)$.

In viscous hydrodynamics the spatial stress tensor  receives derivative corrections
\st
 T^{ij} = \tideal^{ij}(\beta) + \Pi^{ij} \, .
\stp
In the Density Frame the algebraic relations in \Eq{eq:densityframedef} define $\beta_{\mu}(x)$ to all orders in the derivative expansion.
The viscous strains in Bjorken coordinates are proportional to the spatial covariant derivatives of $\beta_{\mu}$:
\st
\label{eq:covderiv}
\Pi^{ij} = -T\kappa^{ijmn} \nabla_{m} \beta_n \equiv - T\kappa^{ijmn} \left( \partial_m \beta_n - \beta_{\tau} \Gamma^{\tau}_{mn} \right) \, .
\stp
A general expression for the tensor $\kappa^{ijmn}(v)$ is given in our companion paper and depends on the shear and bulk viscosities, the flow velocity of the fluid $v^i = \beta^i/\beta^{\tau}$, and the speed of sound~\cite{TheoryPaper}.  In  $1+1$ dimensions with only $x$ dependence,  the symmetries of the problem imply that the only relevant components of 
  $T\kappa^{ijmn}$ are 

\begin{align}
\label{eq:noisekernelintro}
\noisekernel^{xxxx} =&     \frac{1}{(1 -c_s^2 v^2)^2 \gamma^4}   \,  \left( \frac{4\eta}{3}  + \zeta \right)    \,, \\
\tau^2\noisekernel^{xx\eta\eta} =&   \frac{1}{(1 - c_s^2 v^2)^2 \gamma^2}  \left(   -  \frac{2\eta}{3} (1 - 3 c_s^2 v^2)  +  \zeta \right) \,, \\
\tau^4\noisekernel^{\eta\eta\eta\eta} =&   \frac{1}{(1- c_s^2 v^2)^2} \,   \left( \eta (1 - c_s^2 v^2)^2 + \frac{\eta}{3} (1 - 3 c_s^2 v^2)^2  + \zeta  \right)\, .
\end{align}
Here $\gamma =1/\sqrt{1-v^2}$  and $c_s^2 = dp/de$. 
The determinant of this matrix is
\st
\noisekernel^{xxxx}\, \tau^4\noisekernel^{\eta\eta\eta\eta} - (\tau^2\noisekernel^{xx\eta\eta})^2
= \frac{1}{(1 - c_s^2 v^2)^2 \gamma^4}\, \frac{4\eta}{3}\left(\eta + 3\zeta\right) ,
\stp
which is positive for $\eta>0$. For bulk viscosity alone the matrix has rank one, since bulk
viscosity couples only to the expansion rate, and it is positive semi-definite.

The Density Frame and the more familiar Landau frame describe the same stress tensor, but split
it differently into ideal and viscous parts. The Landau-frame energy density $e_L$ and flow
velocity $u_L^\mu$ are defined by the eigenvalue problem
\st
\label{eq:landaumatch}
T^{\mu}{}_{\nu}\, u_L^\nu = -e_L\, u_L^\mu \, ,
\stp
and the Landau-frame viscous stress, $\pi_L^{\mu\nu} = T^{\mu\nu} - (e_L+p_L)u_L^\mu u_L^\nu - p_L g^{\mu\nu}$,
is transverse to $u_L^\mu$. The two frames differ at first order in gradients, so the
Landau-frame temperature and flow velocity differ from the Density Frame $\beta_\mu$ by
first-order corrections. When comparing with kinetic theory (\Sect{sec:kinetic}) and with MUSIC
(\Sect{sec:music}), we construct the full $T^{\mu\nu}$ from the Density Frame solution and
transform it to the Landau frame. Existing particlization codes also assume Landau-frame
variables, so this transformation should be made before freezeout.

\subsection{Entropy production and the dissipative function}
\label{sec:entropy}

Before describing the numerical method, we show that the Density Frame equations produce
entropy, and identify the quadratic form that controls this production, known as the dissipative
function. This quadratic form is the central ingredient of the numerical method in
\Sect{sec:numerics}. Using the equations of motion, it is easy to evaluate the entropy production. 
The entropy density on a constant $\tau$ slice  is defined as in ideal hydrodynamics,  $S
= s(\beta) u^\tau$,  where the intermediate parameters $\beta_{\mu}(\edense,
\pdense^x)$ are functions of the conserved charges. For a fluid cell of volume
$V = \tau\,\Delta x\,\Delta y\,\Delta\eta$, with entropy $\mathcal S = V S$ and four-momentum
$P^\mu = V T^{\tau\mu}$, the thermodynamic identity reads~\cite{TheoryPaper}
\st
d\mathcal S = -\beta_\mu\, dP^\mu + p\,\beta^\tau dV \, .
\stp
Since $V\propto\tau$, the entropy density evolves as
\st
 \partial_\tau (\tau S) = -\beta_{\tau} \partial_\tau (\tau \edense)  -\beta_x  \partial_\tau(\tau \pdense^x) + p\,\beta^\tau \, , 
\stp
and upon using the equations of motion, we find
\st
\partial_\tau( \tau S)  + \partial_x (\tau S v^x -  \beta_x \, \tau \Pi^{xx}) =  \tau 
\begin{pmatrix}
  \partial_x \beta_x   &   -\beta_\tau \Gamma^{\tau}_{\eta\eta}
\end{pmatrix}
\begin{pmatrix}
  T\kappa^{xxxx}  &  T \kappa^{xx\eta\eta} \\
  T \kappa^{\eta\eta xx} & T  \kappa^{\eta\eta\eta\eta} 
\end{pmatrix}
\begin{pmatrix}
  \partial_x \beta_x   \\    -\beta_\tau \Gamma^{\tau}_{\eta\eta}   
\end{pmatrix} \, .
\stp
The right hand side is positive semi-definite as long as the matrix of transport coefficients is positive definite.

We will define the dissipative function $R$ as half the positive semi-definite quadratic form that appears in the entropy production:
\begin{equation}
R[T\kappa^{ijlm}, \beta_i, \beta^\tau] =  \frac{1}{2} \int  \tau  \, \dd x  \, \nabla_{i}\beta_{j} \, \left[  T\kappa^{ijmn}  \right] \,  \nabla_{m}\beta_{n} \, . 
\end{equation}
Regarding the $T\kappa^{ijmn}$ as a constant metric on the space of strains $\nabla_i \beta_j$,  the viscous strains are given by the functional derivative of $R$ with respect to $\nabla_i \beta_j$:
\st
\Gamma^{\tau}_{\eta\eta} \Pi^{\eta\eta} =  \frac{1}{\tau} \frac{\delta R}{\delta
\beta_{\tau}(x)}   \qquad 
\partial_x \Pi^{xx} =  \frac{1}{\tau} \frac{\delta R}{\delta \beta_x(x)} \, .
\stp
When evaluating the rate of entropy production, i.e.  
\begin{equation}
\frac{d\mathbb{S}}{d\tau} \equiv \frac{d}{d\tau} \int \tau \dd x\, S(\tau, x)\, , 
\end{equation}
upon using the equations of motion and dropping the entropy flux at the boundaries, the ideal terms do not contribute, and the viscous terms can be written in terms of the dissipative function as
\begin{equation}
 \frac{d\mathbb{S}}{d\tau} = \int \dd x \left[ \, \beta_\tau \,  (\tau \, \Gamma^{\tau}_{\eta\eta} \Pi^{\eta\eta}) + \beta_x \, \partial_x (\tau\Pi^{xx}) \right] = \int \dd x  \left[  \beta_\tau(x) 
\frac{\delta R}{\delta
\beta_{\tau}(x)} +  \beta_x(x) \frac{\delta R}{\delta \beta_x(x)}  \right]
 = 2 R \, .
\end{equation}
When devising a numerical method,  we will write down the discrete version of the dissipative function and implement the discretized viscous strains by taking derivatives of this function.

\subsection{Numerical method}
\label{sec:numerics}

As in our previous work we will use operator splitting together with Implicit-Explicit (IMEX) time integrators~\cite{ASCHER1997151,Pareschi2005,Constantinescu} to
evolve the system.
The variables
are denoted generically\footnote{For clarity, we will drop the directional superscript in the $1+1$D case,  $\pdense \equiv \pdense^x$.  } $V \equiv(\edense, \pdense)$ and evolve schematically as
\st
  \dot V = f(V) + g(V)\,,
\stp
where $f(V)$ and $g(V)$ represent the ideal and viscous hydrodynamic evolution respectively.
The ideal evolution $f(V)$ is strongly non-linear and will be treated explicitly, while the viscous evolution $g(V)$ will be 
treated implicitly.

In the explicit step, $\dot V = f(V)$,  the variables are evolved over
a time interval $\Delta \tau$ using the equations of motion:
  \begin{subequations}
  \label{eq:FofV}
\begin{align}
  \partial_\tau \edense + \partial_x \pdense =& - \frac{\edense + p(\beta) }{\tau}\, ,   \\
    \partial_\tau \pdense + \partial_x \tideal^{xx}(\beta) =& - \pdense/\tau \, . 
\end{align}
  \end{subequations}
For the explicit step we use a standard Kurganov-Tadmor (KT) scheme of ideal hydrodynamics~\cite{KTScheme,KTAstro,Schenke:2010nt}.

In the implicit step, $\dot V = g(V)$, the variables are evolved
with the viscous equations of motion,
\begin{subequations}
\label{eq:gofV}
\begin{align}
  \partial_\tau \edense
  &= - \Gamma^{\tau}_{\eta\eta} \Pi^{\eta\eta} \, , \\
  \partial_\tau \pdense
  &= - \partial_x \tvisc^{xx} \, .
\end{align}
\end{subequations}
At first order in $\Delta t$, the implicit update has the
schematic form
\st
\label{eq:gofVschematic}
  V^{n+1} = V^n + \Delta t \, g(V^{n+1}) \, ,
\stp
which in general is non-linear and must be
solved with a Newton solver. Combining the explicit and
implicit steps in sequence leads to a non-linear IMEX scheme.
Higher-order IMEX schemes are implemented in
PETSc~\cite{petsc-web-page}, which we use both for time stepping and
for solving the implicit non-linear equations, as in our previous
work~\cite{Bhambure:2024axa}. Our remaining task is to specify in detail the
function $g(V)$ and its Jacobian
$J = \partial g / \partial V$ for the Newton solver.

To this end, we write down the discrete version of the dissipative function $R$. Then   
 taking derivatives of this function with respect to the variables at the $i$-th lattice site, $\beta_{x, i}$ and
$\beta_{\tau, i}$,  defines the discretization.  We will call the vertex located
halfway between the $i$ and $i{+}1$ lattice sites the $i+$ vertex, with an
analogous terminology for the $i-$ vertex. 
A generic variable $u$ at the $i+$ and $i-$ vertices is denoted  $u_{i+}$ and $u_{i-}$,  respectively.  These variables will be defined
as suitable averages  of the variables at the neighboring lattice sites (see
below). 
  We then introduce the \emph{discrete} dissipative function
  \st
  R(T\kappa, \beta) \equiv \tau\Delta x\sum_{i} R_{i+}  \, , 
  \stp 
  where $R_{i+}$ is the contribution to the dissipative function from the $i+$ vertex
  \begin{equation}
  R_{i+} \equiv \frac{1}{2}
\begin{pmatrix}
  \frac{\beta_{x, i+1} - \beta_{x,i}}{\Delta x}  &   -\Gamma^{\tau}_{\eta\eta} \frac{\beta_{\tau, i+1} + \beta_{\tau, i} }{2}
\end{pmatrix}
 \begin{pmatrix}
   T\kappa^{xxxx}   &  T \kappa^{xx\eta\eta} \\
   T\kappa^{\eta\eta xx}   &  T\kappa^{\eta\eta\eta\eta}
 \end{pmatrix}_{i+}
\begin{pmatrix}
  \frac{\beta_{x, i+1} - \beta_{x, i}}{\Delta x}  \\[0.5em]   -\Gamma^{\tau}_{\eta\eta} \frac{\beta_{\tau, i+1} + \beta_{\tau, i} }{2}
\end{pmatrix} \, .
\end{equation}
Then  the discretized viscous strains are given by
\st
(\Gamma^{\tau}_{\eta\eta} \tvisc^{\eta\eta})_i \equiv  
\frac{1}{\tau\Delta x}\frac{\partial R }{\partial \beta_{\tau, i}}  \, , 
\stp
and
\st
(\partial_x \tvisc^{xx})_i\equiv  \frac{1}{\tau\Delta x}\frac{\partial R }{\partial \beta_{x, i}} \, , 
\stp
respectively, which gives  concrete meaning to \cref{eq:gofV}. 
As in the continuum, $R$ is a function of $T\kappa$ and $\beta_{x}$ and
$\beta_{\tau}$ and the  derivatives of $R(T\kappa, \beta)$ with respect to $\beta_{x, i}$ and
$\beta_{\tau, i}$ are to be evaluated by treating $\kappa$ as a constant
metric on the space of strains.  In practice we evaluate $T\kappa^{ijlm}$ at $i+$ by 
averaging the conserved charges $(\edense, \pdense)$ of the neighboring lattice site,  using these
averages to define $\beta_{\mu, i+}$,  and finally evaluating
$T\kappa^{ijlm}(\beta_{i+})$.

There are several benefits to defining the discretization in this way.  First we
note that  the difference structure of the conservation equations is
automatically implemented by the definition of $R$. For example,  the viscous
contribution to the momentum equation is given by the difference of the viscous
stress at the neighboring lattice sites
\st
(\partial_x\Pi^{xx})_i = \frac{\partial R_{i+} }{\partial \beta_{x, i}} + \frac{\partial R_{i-} }{\partial \beta_{x, i}}  
 = 
  \frac{\partial R_{i+} }{\partial \beta_{x, i}} - \frac{\partial R_{i-} }{\partial \beta_{x, i-1}} \equiv  
 \frac{1}{\Delta x} \left(\Pi^{xx}_{i+} - \Pi^{xx}_{i-} \right) \, . 
\stp

Further, with this definition of $R$ the entropy production in the viscous step is automatically
positive semi-definite in a semi-discrete sense. To see this, we note that in the viscous step
only $\edense_i$ and $\pdense_i$ change, at fixed $\tau$, so the total rate of entropy production
on the lattice, with $\mathbb{S} = \sum_i \tau\Delta x\, S_i$, is given by
\begin{align}
  \left.\frac{d\mathbb{S}}{d\tau}\right|_{\rm visc} =
\tau\Delta x \sum_i \left[-\beta_{\tau, i} \,  \partial_\tau \edense_i   - \beta_{x, i} \,   \partial_\tau \pdense_i\right] \, , 
\end{align}
which follows from the definition of the entropy.
Using the semi-discrete viscous equations of motion, \cref{eq:gofV}, the right hand side can be written as
\st
\left.\frac{d\mathbb{S}}{d\tau}\right|_{\rm visc} = 
\tau \Delta x \sum_i \beta_{\tau, i}  (\Gamma^{\tau}_{\eta\eta} \Pi^{\eta\eta})_i + \beta_{x, i} \,
(\partial_x  \Pi^{xx})_i =   \sum_i \beta_{\tau, i} \frac{\partial R}{\partial \beta_{\tau, i}}
+ \beta_{x, i} \frac{\partial R}{\partial\beta_{x, i}} = 2 R  \, .
\stp
Thus the total rate of entropy production in the viscous step is given by $2R$, as in the
continuum.

Now we will write down the explicit form of the viscous strains by differentiating $R$ and introducing a compact notation for the finite differences.
Let $\sigma=0,1$ denote the relative position with respect to the $i$-th lattice site. $\beta_x(\sigma)$  means $\beta_{x, i}$ for $\sigma = 0$ and $\beta_{x, i+1}$ for $\sigma = 1$.  We then introduce the following notation for the finite difference operators
\begin{equation}
  \Kd(\sigma) \equiv   \left(-\frac{1}{2} \Gamma^{\tau}_{\eta\eta}, -\frac{1}{2} \Gamma^{\tau}_{\eta\eta} \right) \, , 
   \qquad D_x(\sigma) \equiv  \left(-\frac{1}{\Delta x}, \frac{1}{\Delta x} \right)  \, , 
\end{equation}
for $\sigma = 0$ and $\sigma = 1$ respectively.
Then the form of $R_{i+}$ is given by 
  \begin{multline}
  R_{i+} = \frac{1}{2}  D_x(\sigma) \beta_x(\sigma) \, T\kappa^{xxxx} \, D_x(\sigma') \beta_x(\sigma') 
   +   \Kd(\sigma) \beta_\tau(\sigma) \, T\kappa^{\eta\eta xx} \, D_x(\sigma') \beta_x(\sigma')   \\
    + \frac{1}{2}  \Kd(\sigma) \beta_\tau(\sigma) \, T\kappa^{\eta\eta\eta\eta} \, \Kd(\sigma') \beta_\tau(\sigma') \, , 
  \end{multline}
where we are summing over the repeated indices $\sigma$ and $\sigma'$. Differentiating with respect to $\beta_{\tau}(\sigma)$ and $\beta_{x}(\sigma)$ yields the appropriate finite difference contribution to the viscous strains at the $i$-th and $(i+1)$-th lattice sites from $R_{i+}$. Thus, the discrete equation 
for the time evolution is  given in part by
\begin{subequations}
  \label{eq:gofVdetail}
\begin{align}
    -\partial_\tau \edense_{i+\sigma}  \supset \frac{\partial R_{i+}}{\partial \beta_\tau(\sigma)} =&  \,   
    \Kd(\sigma) \, T\kappa^{\eta\eta\eta\eta} \, \Kd(\sigma')\beta_\tau(\sigma') 
     + \Kd(\sigma) \, T\kappa^{\eta\eta xx} \, D_x(\sigma')\beta_x(\sigma')  \, , \\
   -\partial_\tau \pdense_{i+\sigma} \supset \frac{\partial R_{i+}}{\partial \beta_x(\sigma)} =& \,   
     D_x(\sigma) T\kappa^{xx \eta\eta} \Kd(\sigma')\beta_\tau(\sigma') + D_x(\sigma) T\kappa^{xxxx} D_{x} (\sigma') \beta_x(\sigma') \, .
\end{align}
\end{subequations}
The complete time evolution is found by summing over the contributions from all  vertices
(i.e. $i+$ and $i-$ for lattice site $i$) to yield 
$(\partial_x \Pi^{xx})_i$ in its complete form. To summarize,
\cref{eq:gofVdetail} defines the function $g(V)$ in the implicit equation
$\partial_\tau V = g(V)$.

The procedure is now easily extended to the 2+1D case.  The two-dimensional 
lattice sites are labeled by the vector $i=(i_x, i_y)$  
and the relative positions are labeled by the vector $\sigma = (\sigma_x, \sigma_y)$ where $\sigma_x$ and $\sigma_y$ can take the values $0$ and $1$.  
$\beta_x(\sigma)$ means $\beta_{x, i+\sigma}$ for 
$\sigma \in \{ \sigma_0, \sigma_1, \sigma_2, \sigma_3\}$ where $\sigma_0 = (0,0)$, $\sigma_1 = (1,0)$, $\sigma_2 = (0,1)$, and $\sigma_3 = (1,1)$.  
We will write down the contribution of $R_{i+(1/2, 1/2)}$ to the viscous strains. 
 The finite difference operators are defined analogously to the 1+1D case,
\begin{subequations}
\begin{align}
  D_x(\sigma)\beta_a(\sigma) \equiv & \frac{1}{2\Delta x} (-\beta_{a, i+\sigma_0} + \beta_{a, i+\sigma_1} - \beta_{a, i+\sigma_2} + \beta_{a, i+\sigma_3}) \, ,  \\
  D_y(\sigma)\beta_a(\sigma) \equiv & \frac{1}{2\Delta y} (-\beta_{a, i+\sigma_0} - \beta_{a, i+\sigma_1} + \beta_{a, i+\sigma_2} + \beta_{a, i+\sigma_3}) \, , \\ 
   \Kd(\sigma) \beta_\tau(\sigma) \equiv & -\Gamma^\tau_{\eta\eta} \frac{1}{4} (\beta_{\tau, i+\sigma_0} + \beta_{\tau, i+\sigma_1} + \beta_{\tau, i+\sigma_2} + \beta_{\tau, i+\sigma_3}) \, .
\end{align}
\end{subequations}
The equations of motion are decorated with transverse indices 
\begin{subequations}
\label{eq:gofVdetail2d}
\begin{align}
     -\partial_\tau \edense_{i+\sigma}  \supset \frac{\partial R_{i+(1/2, 1/2)}}{\partial \beta_\tau(\sigma)} =&  \,   
    \Kd(\sigma) \, T\kappa^{\eta\eta\eta\eta} \, \Kd(\sigma')\beta_\tau(\sigma') 
     + \Kd(\sigma) \, T\kappa^{\eta\eta ab} \, D_a(\sigma')\beta_b(\sigma')  \, ,  \\
   -\partial_\tau \pdense_{i+\sigma}^a \supset \frac{\partial R_{i+(1/2, 1/2)}}{\partial \beta_a(\sigma)} =& \,   
     D_b(\sigma) T\kappa^{a b \eta\eta} \Kd(\sigma')\beta_\tau(\sigma') + D_b(\sigma) T\kappa^{abcd} D_{c} (\sigma') \beta_d(\sigma') \, .
\end{align}
\end{subequations}

To write down an approximate Jacobian for implicit solver, we have to evaluate the
derivatives of the right-hand side of \cref{eq:gofVdetail2d} with respect to 
the variables on the lattice $V_i \equiv T^{\tau \mu}_i \equiv (\edense_i, \pdense_i)$. 
In keeping with the spirit of the dissipative function, we will treat
$T\kappa$ as
a constant, when taking these derivatives.
The Jacobian of the discretized
operator then involves the 
thermodynamic susceptibility  defined by
\begin{align}
\chi^{-1}_{\mu\nu} =&  \frac{\partial \beta_{\mu}}{\partial T^{\tau \nu}}  \, . 
\end{align}
 Explicit formulas for this susceptibility matrix are given in our previous work~\cite{Bhambure:2024axa,TheoryPaper}.
 Using a compact notation  \cref{eq:gofVdetail}  can be written as
\begin{align}
 \partial_\tau T^{\tau \mu}_{i+\sigma}  \supset& \;  g^{\mu\rho}_{\sigma\sigma'}\, \beta_{\rho}(\sigma')  \, , 
\end{align}
and thus the  contribution of $R_{i+(1/2, 1/2)}$ to the Jacobian at sites $i+\sigma$ and $i+ \sigma'$ is 
\begin{align}
 (J_{\sigma, \sigma'})^{\mu}_{\nu} = g^{\mu\rho}_{\sigma\sigma'} \, \chi^{-1}_{\rho\nu}(\sigma')  \qquad (\mbox{no sum over $\sigma'$}). 
\end{align}
Adding the contributions of this sort from all of the vertices,  determines the full Jacobian  matrix for the Newton solver, $J(V) = \partial g/\partial V$.

So far we  have presented a fully non-linear IMEX scheme. A simpler
alternative to a fully non-linear implicit step, known as Extrapolated IMEX (EIMEX), is to linearize
the problem over a time interval $\Delta t$~\cite{Constantinescu}.
We adopted this approach in our previous numerical work on the
Density Frame~\cite{Bhambure:2024axa} and found it to be as
effective and robust as the fully non-linear method, while being
easier to implement and faster too.

In this approach, we solve the linearized equation
\st
\partial_t \, \delta V(t)
= f(V^n) + g(V^n) + J(V^n)\,\delta V \,,
\stp
where $\delta V(t) = V(t) - V^n$ and
$J(V^n) = \partial g/\partial V$ is a time-independent
approximate Jacobian matrix written down in the preceding paragraphs.
This leads to the update
\st
V^{n+1}
= V^n
+ \frac{\Delta t \, (f(V^n) + g(V^n))}
       {1 - \Delta t J(V^n)} \,,
\stp
so that the increment $\delta V$ is obtained by solving the
linear system
\st
(1 - \Delta t J(V^n)) \, \delta V
= \Delta t \, (f(V^n) + g(V^n)) \,.
\stp
The first order EIMEX scheme presented here,  as well as its higher-order generalizations, are all implemented in
PETSc~\cite{petsc-web-page}.  We adopted a third-order EIMEX scheme studied in
our previous work~\cite{Bhambure:2024axa}.
We used PETSc both for EIMEX time
stepping and to solve the linear systems arising in the
linearized treatment. In practice, the default linear solver in
PETSc---a GMRES solver~\cite{trefethen1997numerical} with either an
ILU or a Jacobi preconditioner---has proven to be efficient for these
linear systems.

\section{Comparison to the Gubser solution: an exact kinetic benchmark}
\label{sec:kinetic}

To test Density Frame hydrodynamics we need a reference solution with transverse expansion
in which the approach to hydrodynamics is under control. Gubser flow provides one. It is a
conformal, boost-invariant and azimuthally symmetric flow with both longitudinal and radial
expansion~\cite{Gubser:2010ze,Gubser:2010ui}, and the Boltzmann equation in the relaxation
time approximation (RTA) can be solved exactly with the same symmetry~\cite{Denicol:2014xca,
Denicol:2014tha}. The kinetic solution and its relation to hydrodynamics have been studied
extensively~\cite{Chattopadhyay2018gubser,Strickland2016aHydroGubser,Dash:2020zqx}. However, how
the kinetic solution matches onto hydrodynamics has not been fully explained. Here we present
it in a form suited to heavy-ion collisions, with the initial conditions set at $\tau\to0^+$,
and organize it by the opacity~\cite{Kurkela2019flow,Ambrus2022range,Ambrus2022opacity,Arslandok:2023utm,Schlichting:2019abc}.
We show how the Bjorken
attractor~\cite{Heller2015attractor,Strickland2018rta,Giacalone2019attractor,Schlichting:2019abc} passes onto the
Gubser solution order by order in the gradient expansion, and use this to make a clean
comparison with Density Frame hydrodynamics.

Gubser's hydrodynamic solution has a single integration constant, which we take to be the
\emph{central temperature} $T_0$, i.e. the temperature at $r=0$ and $\tau=R$. This integration constant fixes the entire Gubser flow 
to all orders in the gradient expansion~\cite{DenicolNoronha2018slowroll}. Our goal in this
section is to relate $T_0$ to the initial conditions. 
The initial conditions have a momentum
scale $Q_1$ and a transverse size $R$,\footnote{$R = 1/q$ in the notation of
Ref.~\cite{Gubser:2010ze} and most of the subsequent literature.}  and the dynamics is set by one transport coefficient, $\eta/s$.
We work with kinetic theory, which is valid in the semi-classical limit where $Q_1R/\hbar\gg1$
and $\eta/(s\hbar)\gg1$, i.e.\ momenta are large compared with $\hbar/R$ and the mean free path is
long compared with the thermal wavelength.\footnote{Here and below we occasionally restore
$\hbar$ to make the semi-classical limit transparent.} In this limit, with
momenta measured in units of $Q_1$ and lengths and times in units of $R$, the solution of the
kinetic equation depends on $Q_1$, $R$ and $\eta/s$ only through the opacity
\begin{equation}
  \chi_1 \equiv \frac{(Q_1R/\hbar)^{2/3}}{\bigl(5\eta/(s\hbar)\bigr)^{8/9}}\, ,
  \label{eq:chi_orig}
\end{equation}
which measures the number of relaxation times a particle experiences over a time $R$ (see below). The semi-classical limit is taken with $\chi_1$ held fixed, i.e.\ $Q_1R/\hbar\to\infty$ and
$\eta/(s\hbar)\to\infty$ with $\eta/(s\hbar)\propto(Q_1R/\hbar)^{3/4}$.
We determine $T_0$ for all opacities, from free streaming to the hydrodynamic regime
(\Fig{fig:Trhozero}). Since $T_0$ fixes the hydrodynamic flow, this gives a complete prediction
for Gubser flow in terms of the initial conditions and $\eta/s$, and the full kinetic solution
then provides a clean test of the Density Frame (\Sect{sec:df}).

This section uses Milne coordinates $(\tau,r,\phi,\varsigma)$, with
$\varsigma$ the spacetime rapidity; the Gubser (de~Sitter) technology is collected in
\app{app:gubser}.

\subsection{Kinetic theory and initial conditions}

We solve the conformal Boltzmann equation in RTA,
\begin{equation}
  p^\mu\partial_\mu f = \frac{u\cdot p}{\tau_R}\,\bigl(f - f_{\rm eq}\bigr), \qquad
  \tau_R = \frac{\cRTA}{T_{\rm eff}} \, ,
\end{equation}
where $u^\mu$ is the Landau velocity and $T_{\rm eff}$ is defined by the energy density,
$\varepsilon = \Ce T_{\rm eff}^4$ with $\Ce = 3/\pi^2$. The constant $\cRTA$ is related to the
shear viscosity by $\cRTA = 5\eta/s$~\cite{Denicol:2014tha}. Classical kinetic theory requires
$T_{\rm eff}\tau_R = 5\eta/s\gg1$, i.e.\ weak coupling; otherwise the system would relax faster than $\hbar/T_{\rm eff}$.

At $\tau = 0^+$ the particles are produced with zero longitudinal momentum in the local frame
($p_\varsigma = 0$) and with a transverse-momentum spectrum and density that depend on $r$:
\begin{equation}
  (2\pi)^3\frac{\dd N}{\dd^2p_T\,\dd p_\varsigma\,\dd^2r\,\dd\varsigma}
  = \Ce\Kinf\,\frac{4p_T}{3Q(r)}K_1\!\left(\frac{p_T}{Q(r)}\right)2\pi\delta(p_\varsigma),
  \qquad Q(r) = \frac{2Q_1}{1+(r/R)^2}\, ,
  \label{eq:yield}
\end{equation}
where $K_1$ is a modified Bessel function. The initial energy density is
$\tau\varepsilon = \Ce\Kinf Q^3(r)$. The order-one constant $\Kinf$ fixes the
normalization of $Q_1$, and is determined below by requiring that the final entropy be simply
related to $Q_1$ in the hydrodynamic regime.
This profile makes the problem exactly solvable: it arises from a thermal distribution in
Gubser's de~Sitter coordinates at $\rho\to-\infty$, which then streams freely (\app{app:ic}).
It is also similar in form to the initial parton distributions of the Color Glass
Condensate~\cite{Gelis:2010nm,Schlichting:2019abc}, with $Q_1$ playing the role of the saturation scale. With these initial
conditions the kinetic equation reduces to an ordinary differential equation in one variable,
which we solve with the moment method of Ref.~\cite{Dash:2020zqx} (\app{app:kinetic}).

\subsection{From free streaming to Gubser flow}
\label{sec:timeline}

Consider first a system that hydrodynamizes at a time $\tau_H\ll R$. The process
is a local one and happens independently at each point in the transverse plane. The effective temperature has three evolution 
stages~\cite{Arslandok:2023utm}, each with a constant of motion:
\begin{enumerate}[label=(\alph*)]
\item For $\tau\ll\tau_H$ the particles stream freely along the beam axis at each
  transverse point. The energy per rapidity and area is constant, and the effective temperature falls
  as $\tau^{-1/4}$:
  \begin{equation}
    \tau\varepsilon(r) = \Ce\Kinf\,Q^3(r)\, , \qquad
    T_{\rm free}(\tau,r) \equiv \Kinf^{1/4}\left(\frac{Q^3(r)}{\tau}\right)^{1/4}.
  \end{equation}
\item For $\tau_H\ll\tau\ll R$ the system is a nearly ideal fluid undergoing Bjorken
  expansion at each transverse point. The entropy per rapidity and area is constant, and with $s\propto T^3$,
  \begin{equation}
    \tau T^3 = \Lambda^2(r)\, , \qquad
    T_{\rm bj}(\tau,r) \equiv \left(\frac{\Lambda^2(r)}{\tau}\right)^{1/3}.
  \end{equation}
\item For $\tau\gtrsim R$ the radial expansion starts, and the fluid follows Gubser's
  ideal solution,
  \begin{equation}
    T(\tau,r) = \frac{\hat T_0}{\tau\cosh^{2/3}\rho}, \qquad v^r = \frac{2\tau r}{R^2+\tau^2+r^2}\, ,
    \qquad
    \sinh\rho = -\frac{R^2-\tau^2+r^2}{2\tau R}\, ,
    \label{eq:Tideal}
  \end{equation}
  with a dimensionless constant $\hat T_0$. Here $\rho$ is Gubser's time variable;
  $\rho\to-\infty$ at early times and $\rho=0$ on the surface $\tau^2 = R^2+r^2$, so that
  $\hat T_0 = R\,T_0$, with $T_0$ the central temperature defined below.
\end{enumerate}
Our task is to relate the three constants $Q_1\equiv Q(R)$, $\Lambda_1 \equiv \Lambda(R)$ and $\hat T_0$. Here and below, the subscript $1$ refers to $r=R$. Similarly, $T_{\rm free}(\tau) \equiv T_{\rm free}(\tau, R)$ and $T_{\rm bj}(\tau)\equiv T_{\rm bj}(\tau, R)$, written with one argument, refer to $r=R$.

We will study the \emph{central temperature}, defined as the temperature at the center of the fireball at $\tau=R$,
\begin{align}
T_{0} \equiv  \left. T_{\rm eff}(\tau, r=0) \right|_{\tau=R}\, .
\end{align}
In the hydrodynamic regime, where the Gubser solution applies, $T_{0} = \hat T_0/R$.  The central temperature can be estimated by extrapolating either the free result or the Bjorken flow result to time  $\tau=R$. This leads us to define two \emph{estimates}  for the central temperature,
\begin{align} 
T_{\rm free}(R) \equiv \Kinf^{1/4} \left( \frac{Q_1^3}{R} \right)^{1/4}
 \qquad  
T_{\rm bj}(R) \equiv  \left( \frac{\Lambda_1^2}{R} \right)^{1/3} .
\end{align}
These estimates are parametrically valid in the free-streaming ($\chi_1\ll1$) and hydrodynamic
($\chi_1\gg1$) regimes, respectively. In fact, these estimates happen to reproduce the central
temperature exactly in the corresponding limits (see below), though in general one would expect a
numerical factor of order one. We will see shortly that for
$\chi_1\ll1$, $T_{\rm free}(R)$ is parametrically smaller than $T_{\rm bj}(R)$. $T_{\rm bj}(R)$ will
serve as our unit of temperature in what follows.

Viscous corrections to ideal Bjorken flow are controlled by the ratio of the relaxation time to the age of the system,
$\tau_R/\tau  = 5\eta/(sT \tau)$. This motivates a definition of the hydrodynamization time,
\begin{equation}
  \left(\frac{\tau_H}{\tau}\right)^{2/3}\equiv\frac{5\eta/s}{\tau\,T_{\rm bj}(\tau)}\, ,
  \qquad \tau_H = \frac{(5\eta/s)^{3/2}}{\Lambda_1}\, .
\end{equation}
For $\tau_H\ll\tau\ll R$, the viscous corrections form a series in $(\tau_H/\tau)^{2/3}$,
\begin{equation}
  T_{\rm eff}(\tau) = T_{\rm bj}(\tau)\left[1 + b_1\left(\frac{\tau_H}{\tau}\right)^{2/3}
  + b_2\left(\frac{\tau_H}{\tau}\right)^{4/3} + \dots\right].
  \label{eq:bj_series}
\end{equation}
The first coefficient, $b_1 = -2/15$, follows from Navier--Stokes. The second depends on the
second-order coefficients through $\eta\tau_\pi - \lambda_1$  (see \eq{eq:b2_general}); for RTA,
$\eta\tau_\pi = 5\,\eta^2/(e+P)$ and $\lambda_1 = \tfrac{25}{7}\,\eta^2/(e+P)$~\cite{TeaneyYan2013}, and so $b_2 = -4/315$.

For times $\tau \sim \tau_{H}$, the temperature follows the Bjorken
attractor~\cite{Heller2015attractor,Strickland2018rta,Giacalone2019attractor,Kurkela2018kompost},
an approximately universal function interpolating between free streaming and the Bjorken solution,
$T_{\rm eff} = T_{\rm bj}(\tau)\,\mathcal T(\tau_H/\tau)$. This interpolating function has the limits
\begin{equation}
  \mathcal T \simeq
  \begin{cases}
    \Kinf^{1/4}\,(\tau/\tau_H)^{1/12}, & \tau\ll\tau_H\, ,\\[4pt]
    1 - \tfrac{2}{15}(\tau_H/\tau)^{2/3}, & \tau\gg\tau_H\, .
  \end{cases}
\end{equation}
We fix $\Kinf$ by requiring that $\mathcal T\to1$ as $\tau\to\infty$. Solving the kinetic
equations numerically, we find\footnote{This is equivalent to
$\Cinf\simeq0.874$ in the notation of Ref.~\cite{Giacalone2019attractor}.} $\Kinf \simeq 0.734$. This effectively defines the
relation between $Q_1$ and $\Lambda_1$,
\begin{equation}
  \Lambda_1 = (5\eta/s)^{1/6}\,Q_1\, .
  \label{eq:Lambda_Q}
\end{equation}
At $\tau = \tau_H$, the two limiting temperatures $T_{\rm free}(\tau_H)$ and $T_{\rm bj}(\tau_H)$
differ only by a factor $\Kinf^{1/4}\simeq0.93$.

The opacity compares the hydrodynamization time $\tau_H$ to the radius, and equals the number
of relaxation times at $\tau = R$:
\begin{equation}
  \chi_1 = \left(\frac{\tau_H}{R}\right)^{-2/3} = \frac{R\,T_{\rm bj}(R)}{5\eta/s}\, .
\end{equation}
Combining \eq{eq:Lambda_Q} with the definition of
$\chi_1$ gives the unintuitive combination of parameters in\footnote{Other definitions of the opacity are in use. The
opacity $\hat\gamma$ of Refs.~\cite{Kurkela2019flow,Ambrus2022range,Ambrus2022opacity},
$\hat\gamma = (5\eta/s)^{-1}\bigl(R\,\dd E_\perp/\dd y/(\pi \Ce)\bigr)^{1/4}$, is related to
ours by $\chi_E = \Kinf^{-2/9}\hat\gamma^{8/9}$ and $\chi_1 = 4^{-2/9}\chi_E$. Here
$\chi_E\equiv(\tau_H/R)_E^{-2/3}$, where $(\tau_H/R)_E^{-3}$ is the average of $(\tau_H(r)/R)^{-3}$
over $\pi R^2$. The opacity of Ref.~\cite{Arslandok:2023utm}, defined with
$4\pi\eta/s$, is $5\chi_E/4\pi$.} \Eq{eq:chi_orig}.
In terms of $\chi_1$, the expansion parameter of \eq{eq:bj_series} is
$(\tau_H/\tau)^{2/3} = \chi_1^{-1}(R/\tau)^{2/3}$, and the Bjorken series reads
\begin{equation}
  T_{\rm eff}(\tau) = T_{\rm bj}(\tau)\left[1 + \frac{b_1}{\chi_1}\left(\frac{R}{\tau}\right)^{2/3}
  + \frac{b_2}{\chi_1^2}\left(\frac{R}{\tau}\right)^{4/3} + \dots\right].
  \label{eq:bj_series_chi}
\end{equation}
At the boundary of applicability, $\tau \lesssim R$, the series is organized in powers of
$1/\chi_1$, which must match the subsequent evolution.

For $\tau\sim R$, the radial expansion matters. Gubser's ideal solution is then the leading
term, and the viscous corrections form an expansion in
$1/\chi$~\cite{Gubser:2010ze,Marrochio2013IS,DenicolNoronha2018slowroll}. Through second order
(\app{app:hydro}),
\begin{equation}
  T(\tau,r) = \frac{\hat T_0}{\tau\cosh^{2/3}\rho}\left[1 + \frac{F^{(1)}(\rho)}{\chi_0}
  + \frac{F^{(2)}(\rho)}{\chi_0^2} + \dots\right],
  \qquad \chi_0 \equiv \frac{R\,T_0}{5\eta/s}\, ,
\end{equation}
with the flow of \eq{eq:Tideal}. Here $F^{(1)}$ is Gubser's first-order correction, a
hypergeometric function~\cite{Gubser:2010ze}, and $F^{(2)}$ is elementary; both are given in
\app{app:hydro} [Eqs.~(\ref{eq:delta1_general}) and (\ref{eq:delta2_general})]. Both vanish at
$\rho = 0$, and, in general, the central temperature is $T_0 = \hat T_0/R$ to all orders in the
hydrodynamic gradient expansion. At early times, this expansion must reproduce
the Bjorken series \eq{eq:bj_series} at every transverse point. For the ideal solution, this
works because $\Lambda(r)$ and $Q(r)$ have the same profile, and the first- and second-order
terms reproduce $b_1$ and $b_2$. The constant $\hat T_0$, however, has to be re-matched at each
order. Matching at $r=R$ gives
\begin{equation}
  \frac{T_0}{T_{\rm bj}(R)} = 1 + \frac{\aone}{\chi_1} + \frac{\atwo}{\chi_1^2}
  + \mathcal O\!\left(\frac{1}{\chi_1^3}\right),
  \label{eq:Thydro_center}
\end{equation}
with
\begin{equation}
  \label{eq:a1a2}
  \aone = -\frac{4\,\Gamma(\tfrac32)\Gamma(\tfrac23)}{5\,\Gamma(\tfrac16)}\simeq-0.1725\, ,
  \qquad \atwo = \frac{2}{75}\,\frac{\lambda_1(e+P)}{\eta^2} = \frac{2}{21}\, .
\end{equation}
Unlike $b_2$, the coefficient $\atwo$ depends only on $\lambda_1$. The opacity parameter $\chi_0$ equals $\chi_1$ only at lowest order in the gradient expansion.

In the opposite limit, $\chi_1\ll1$, the initial condition streams freely through the radial
expansion, giving~\cite{Denicol:2014tha}
\begin{equation}
  \varepsilon(\tau,r) = \frac{\Ce\Kinf\,(Q_1R)^3}{\tau^4\cosh^3\rho}\, , \qquad
  T_{\rm eff}(\tau,r) = \frac{T_{\rm free}(R)\,R}{\tau\cosh^{3/4}\rho}\, .
\end{equation}
Thus, the central temperature in the free regime is parametrically smaller than $T_{\rm bj}(R)$:
\begin{equation}
  \frac{T_0}{T_{\rm bj}(R)} = \frac{T_{\rm free}(R)}{T_{\rm bj}(R)} = \Kinf^{1/4}\chi_1^{1/8}\, .
  \label{eq:Tfree_center}
\end{equation}
Here we have used \eq{eq:Lambda_Q} and the definition of $\chi_1$. Without longitudinal
work~\cite{Danielewicz:1984ww}, the free-streaming temperature falls more slowly than in ideal
hydrodynamics, as $\tau^{-1/4}$ rather than $\tau^{-1/3}$.
For $\chi_1\ll1$ the time $\tau=R$ comes well before $\tau_H$, so extrapolating the hydrodynamic
behavior back to $\tau=R$ overestimates the temperature.

\begin{figure}[t]
  \centering
  \includegraphics[width=0.95\linewidth]{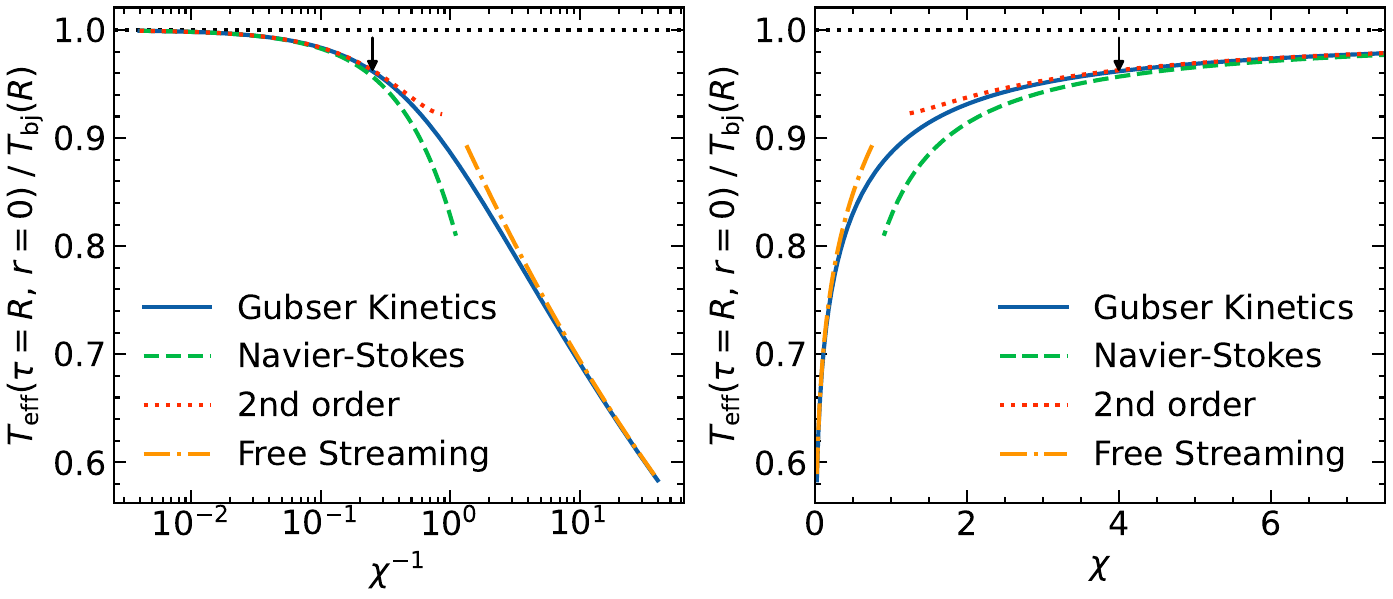}
  \caption{Central temperature $T_0$ at $\tau=R$ in units of $T_{\rm bj}(R)$ versus inverse opacity
  (left) and opacity (right). Solid: RTA kinetic theory. Dashed: first-order hydrodynamics,
  $1+\aone/\chi_1$. Dotted: second-order hydrodynamics, $1+\aone/\chi_1+\atwo/\chi_1^2$, shown up
  to its minimum at $\chi_1^{-1}\simeq0.9$. Dash-dotted: free streaming,
  $\Kinf^{1/4}\chi_1^{1/8}$. The arrows mark $\chi_1=4$, the opacity used in \Fig{fig:gubser_df}.}
  \label{fig:Trhozero}
\end{figure}

\Fig{fig:Trhozero} shows the central temperature from the kinetic solution as a function of
opacity, together with the hydrodynamic expansion \eq{eq:Thydro_center} and the
free-streaming result \eq{eq:Tfree_center}. The kinetic result interpolates smoothly between
the two limits. The figure is therefore a complete prediction for the central temperature $T_0$,
and hence for Gubser flow in the hydrodynamic regime, which is completely fixed by $T_0$.
Specifically, given $Q_1$, $R$ and $\eta/s$, with the initial energy density
$\tau\varepsilon = \Ce\Kinf Q^3(r)$ and $\Kinf\simeq0.734$, the opacity follows from
\eq{eq:chi_orig}, and the figure gives $T_0$ in units of
$T_{\rm bj}(R) = (5\eta/s)^{1/9}\,(Q_1^2/R)^{1/3}$. The opacity alone thus determines how
close the central temperature comes to the hydrodynamic limit.

\subsection{Density frame and the kinetic solution}
\label{sec:df}

In the Density Frame the only evolution variables are the conserved densities
$T^{\tau\tau}$ and $T^{\tau i}$. The Density Frame energy density and velocity are
defined by $T^{\tau\tau} = (e+p)u^\tau u^\tau - p$ and $T^{\tau i} = (e+p)u^\tau u^i$, with no
viscous part in $T^{\tau\mu}$. The viscous stress appears only in the spatial components,
$T^{ij} = (e+p)u^iu^j + p\,g^{ij} + \pi^{ij}$, and $\pi^{ij}$ is fixed at each time by the
constitutive relation. It is not an independent variable~\cite{TheoryPaper}.

To compare with kinetic theory, we construct the full stress tensor from the Density Frame
solution and transform it to the Landau frame, as described at the end of \Sect{sec:eom}
[see \Eq{eq:landaumatch}].

At an initial time $\tau_0$, we take from the kinetic solution only the conserved densities
$T^{\tau\tau}$ and $T^{\tau i}$. These are the Density Frame variables, and everything else
follows from the constitutive relations. As a result, the Density Frame viscous stress at
$\tau_0$ is not the kinetic one: it takes the value set by the gradients of the initial
$T^{\tau\mu}$, while the kinetic stress remembers the earlier evolution. In
Israel--Stewart-type theories, $\pi^{\mu\nu}$ is an independent variable and could be
initialized from kinetic theory; the Density Frame has no such freedom.

This sets the right way to compare. We fix $\tau_0/R$ and increase $\chi$. The mismatch at
$\tau_0$ then shrinks as $\chi$ grows, and the error of the comparison shrinks with it.
If instead $\tau_0/\tau_H$ were held fixed, the system would always be initialized at the same
stage of hydrodynamization, and the error would not decrease. Avoiding this initialization
error is the main reason for comparing with the full kinetic solution, which provides a
consistent initial state at any $\tau_0/R$.

\begin{figure}[t]
  \centering
  \includegraphics[width=0.48\linewidth]{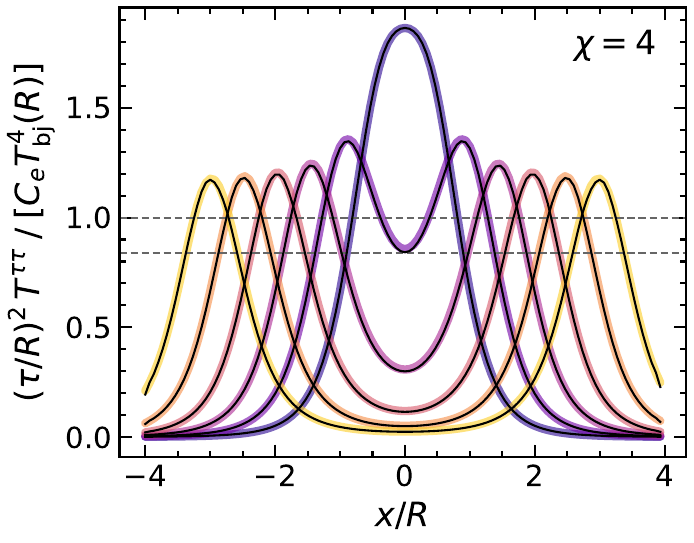}\hfill
  \includegraphics[width=0.48\linewidth]{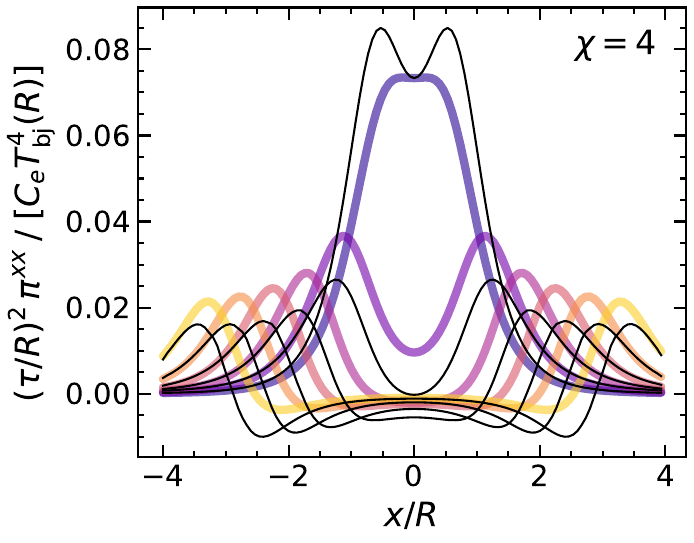}
  \caption{Gubser flow with $\chi = 4$. $y = 0$ slices of the
  energy density $(\tau/R)^2T^{\tau\tau}$ (left) and the Landau-frame shear stress $(\tau/R)^2\pi^{xx}$
  (right), in units of $\Ce T_{\rm bj}^4(R)$, at $\tau/R = 0.5, 1.0, \dots, 3.0$ (dark to light).
  Thick colored lines: RTA kinetic theory. Thin black lines: Density Frame hydrodynamics,
  initialized with the kinetic $T^{\tau\tau}$ and $T^{\tau r}$ at $\tau_0 = 0.5R$ and transformed to
  the Landau frame. Dashed lines (left): the central value at $\tau = R$ for ideal hydrodynamics, $1$, and
  for Navier--Stokes, $(1+\aone/\chi)^4$, corresponding to the temperature shift marked by
  the arrows in \Fig{fig:Trhozero}.}
  \label{fig:gubser_df}
\end{figure}

We initialize at $\tau_0 = 0.5R$ and evolve to $\tau = 3R$ on a $120\times120$ grid covering
$[-4R,4R]^2$. Since $\tau_0/\tau_H = 0.5\,\chi^{3/2}$, $\tau_0 = 4\tau_H$ at $\chi = 4$.
\Fig{fig:gubser_df} compares the two at $\chi = 4$, both shown in the Landau frame. We note
the following.
\begin{enumerate*}[label=(\roman*), itemjoin={{ }}, itemjoin*={{ }}]
\item The differences in the energy density $T^{\tau\tau}$ (left panel) are small, hard to see
  in the figure, and of order $1/\chi^2$.
\item The dashed lines in this panel indicate the central energy density $\Ce T_{\rm eff}^4$ at
  $\tau=R$ in ideal hydrodynamics, where $T_{\rm eff}=T_{\rm bj}(R)$, and the corresponding
  first-order viscous reduction, $(1+\aone/\chi)^4$. This shift reflects the arrows in
  \Fig{fig:Trhozero}.
\item The right panel shows the shear stress, which is smaller than the energy density by a
  factor of order $1/\chi$. Its deviation from the kinetic result is of order $1/\chi^2$
  (i.e.\ $\delta\pi/\pi\sim1/\chi$).%
\end{enumerate*}
We have studied a range of $\chi$, and all
components of the stress tensor along different directions in the transverse plane, with
similar results.

These comparisons also test the approximate foliation independence, and by extension the
Lorentz invariance, of the Density Frame. With $\rho$ as the time variable the fluid is at rest, and the Density Frame
solution is Gubser's hydrodynamic solution. Our code instead uses $\tau$ slices, on which the
fluid flows radially, and it reproduces the kinetic solution, and hence Gubser's hydrodynamic
solution, up to higher-order gradient corrections.

\section{Comparison to MUSIC for typical heavy-ion initial conditions}
\label{sec:music}

\begin{figure}[!ht]
    \centering
    \includegraphics[width=0.9\linewidth]{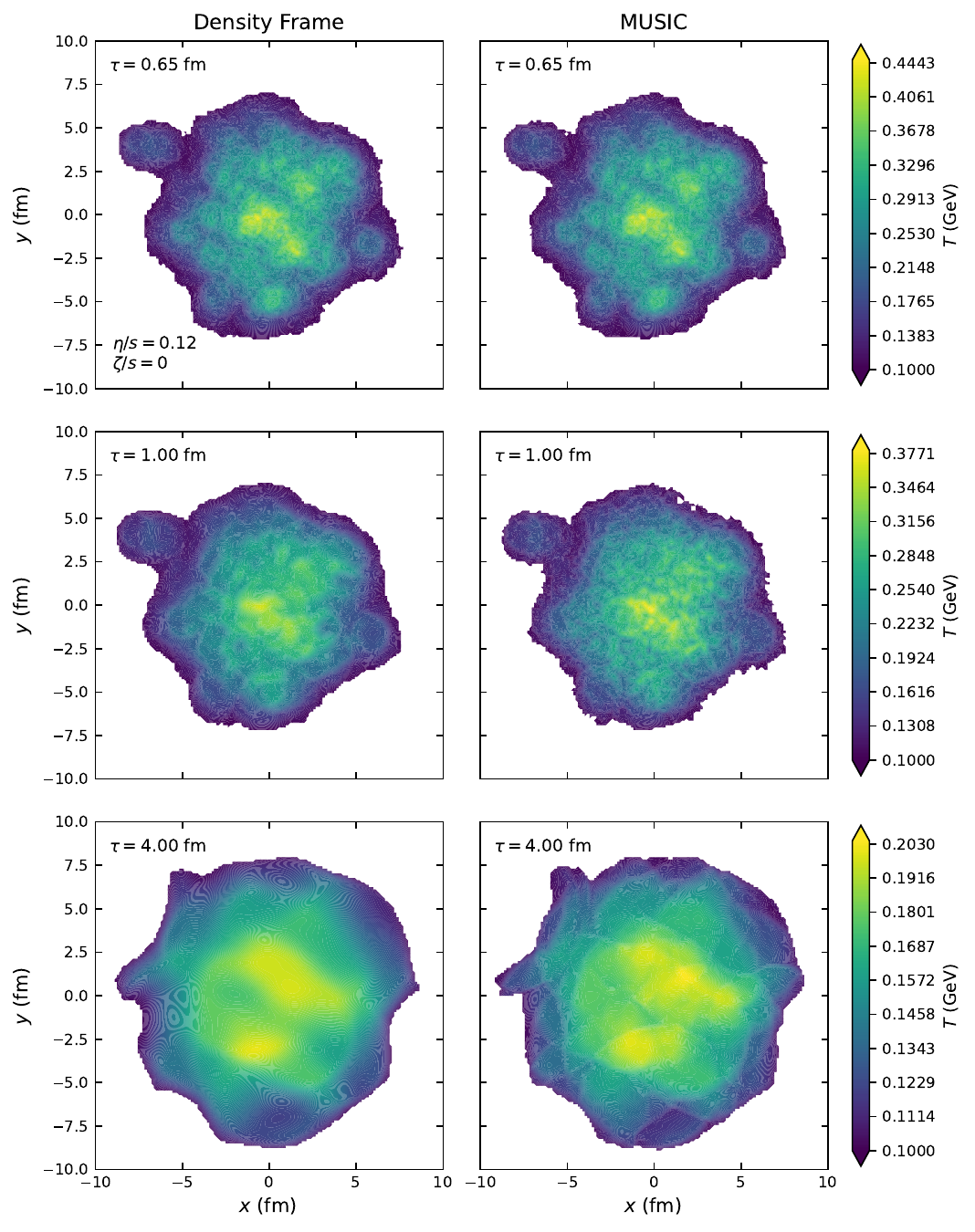}
    \caption{Evolution of the temperature ($T$) profile for a single IP-Glasma event using Density Frame hydrodynamics (left) and MUSIC (right). The temperature is shown at proper time $\tau = $ 0.65 fm (top), 1 fm (middle) and 4 fm (bottom). Both simulations use the same initial temperatures initialized at $\tau_0 = 0.6$ fm with shear viscosity $\eta/s=0.12$ and zero bulk viscosity.}
    \label{fig:temperature_2d}
\end{figure}

We compare the 2+1D Density Frame hydrodynamic evolution with the Israel–Stewart (more specifically, DNMR~\cite{Denicol:2012cn}) implementation in MUSIC \cite{Schenke:2010nt,Schenke:2010rr,Paquet:2015lta} using realistic initial conditions for heavy-ion collisions. Both simulations are initialized with the same IP-Glasma event \cite{Schenke:2012wb,Schenke:2012hg} at a proper time of $\tau_0 = 0.6$ fm and evolved in Milne coordinates. The IP-Glasma + MUSIC framework has been extensively used to model heavy-ion collisions and successfully describe a broad range of experimental observables. For both simulations, we use a QCD equation of state constructed by smoothly matching lattice QCD results at high temperatures \cite{HotQCD:2014kol} to a hadron resonance gas at low temperatures. The shear viscosity-to-entropy density ratio is held constant at $\eta/s = 0.12$, while the bulk viscosity is initially set to zero.

In the second-order viscous hydrodynamic framework employed by MUSIC, the shear-stress tensor and bulk viscous pressure are independent dynamical fields that require explicit initialization. While these quantities were often initialized to zero, the current practice when coupling IP-Glasma to MUSIC is to initialize them by matching the energy-momentum tensor across the transition, thereby ensuring energy-momentum conservation. Specifically, the IP-Glasma energy-momentum tensor is decomposed through Landau matching to extract the local energy density and flow velocity, while its traceless component transverse to the fluid velocity is used to initialize the shear-stress tensor. However, because IP-Glasma is conformal, whereas MUSIC typically uses a nonconformal QCD equation of state, a mismatch arises between their equilibrium pressures. To compensate for this difference while conserving energy and momentum at initialization, the bulk viscous pressure is initialized accordingly, generally resulting in a positive value. Such a positive bulk pressure enhances the effective isotropic pressure and can contribute to the fluid expansion, in contrast to the negative bulk pressure typically associated with an expanding viscous fluid, which acts to resist expansion. There have long been concerns about this bulk pressure initialization, which can have phenomenological implications~\cite{NunesdaSilva:2020bfs}. In Density Frame hydrodynamics, by contrast, the shear-stress tensor and bulk viscous pressure are not independent dynamical variables and therefore do not require separate initialization. Instead, only the energy density and flow velocity are extracted from IP-Glasma and used as the Density Frame energy density and flow velocity,\footnote{Alternatively, one could initialize the Density Frame by identifying the IP-Glasma $T^{\tau\tau}$ and $T^{\tau i}$ directly with $\edense$ and $\pdense^i$, which conserves energy and momentum exactly at the transition; this is what we do in the comparison with Gubser flow (\Sect{sec:df}). We have found that the resulting flow fields are very close in practice. In either case, the IP-Glasma energy-momentum tensor contains more independent degrees of freedom than are required to initialize the Density Frame evolution.} while the viscous contributions are determined by their Navier–Stokes constitutive relations.

\begin{figure}[!ht]
    \centering
    \includegraphics[width=0.9\linewidth]{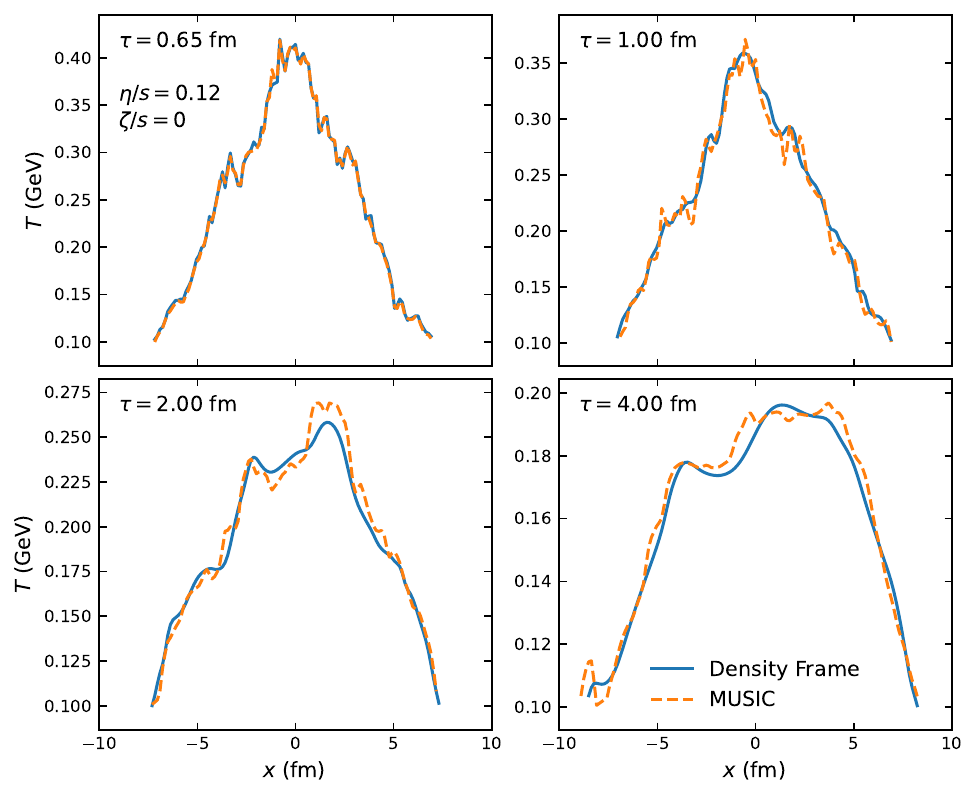}
    \caption{Temperature profiles along the transverse slice ($y=0$) obtained with Density Frame hydrodynamics (solid blue) and MUSIC (dashed orange) at proper times $\tau=$ 0.65, 1.0, 2.0 and 4.0 fm. Both simulations are initialized with the same IP-Glasma event at $\tau_0 = 0.6$ fm and have $\eta/s=0.12$ and $\zeta/s=0$.}
    \label{fig:temperature_1d}
\end{figure}

An important difference between the two implementations lies in their treatment of low-density regions, where numerical noise can generate large, spurious gradients and lead to instabilities. In MUSIC, these instabilities are controlled by applying regulators in such regions. In contrast, the Density Frame implementation requires only a small energy-density pedestal. At the initial time, cells with energy densities below $5\times10^{-4}$ fm$^{-4}$ are assigned this minimum value, with no further regulation applied during the subsequent evolution. 

\subsection{Hydrodynamic fields}
\label{sec:music_hydro_fields}

\begin{figure}[!ht]
    \centering
    \includegraphics[width=0.9\linewidth]{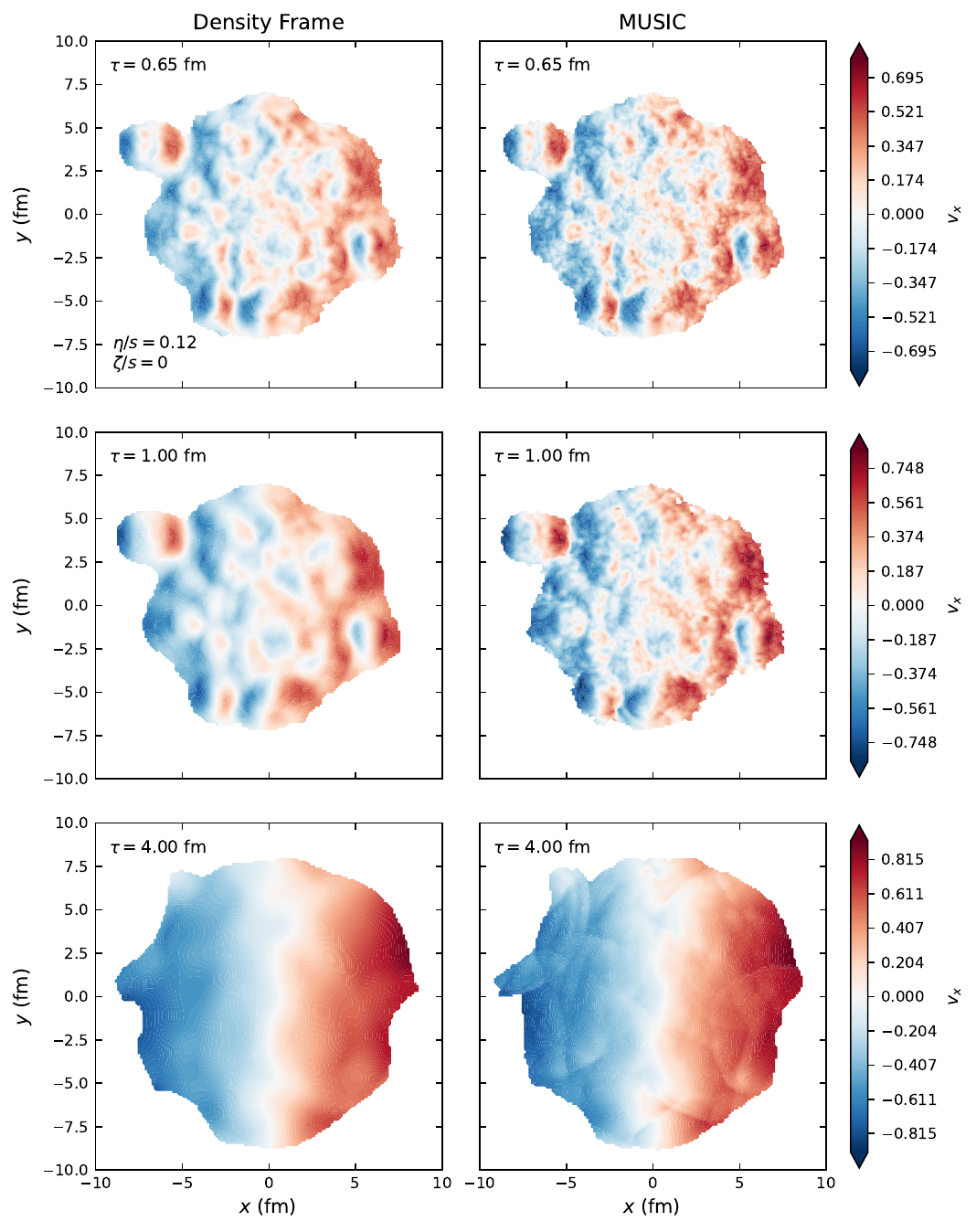}
    \caption{Evolution of the $x-$component of transverse velocity ($v_x$) profile for a single IP-Glasma event using Density Frame hydrodynamics (left) and MUSIC (right). The velocity field is shown at proper time $\tau = $ 0.65 fm (top), 1 fm (middle) and 4 fm (bottom). Both simulations use the same initial velocities initialized at $\tau_0 = 0.6$ fm with shear viscosity $\eta/s=0.12$ and zero bulk viscosity.}
    \label{fig:vx_2d}
\end{figure}

\begin{figure}[!ht]
    \centering
    \includegraphics[width=0.9\linewidth]{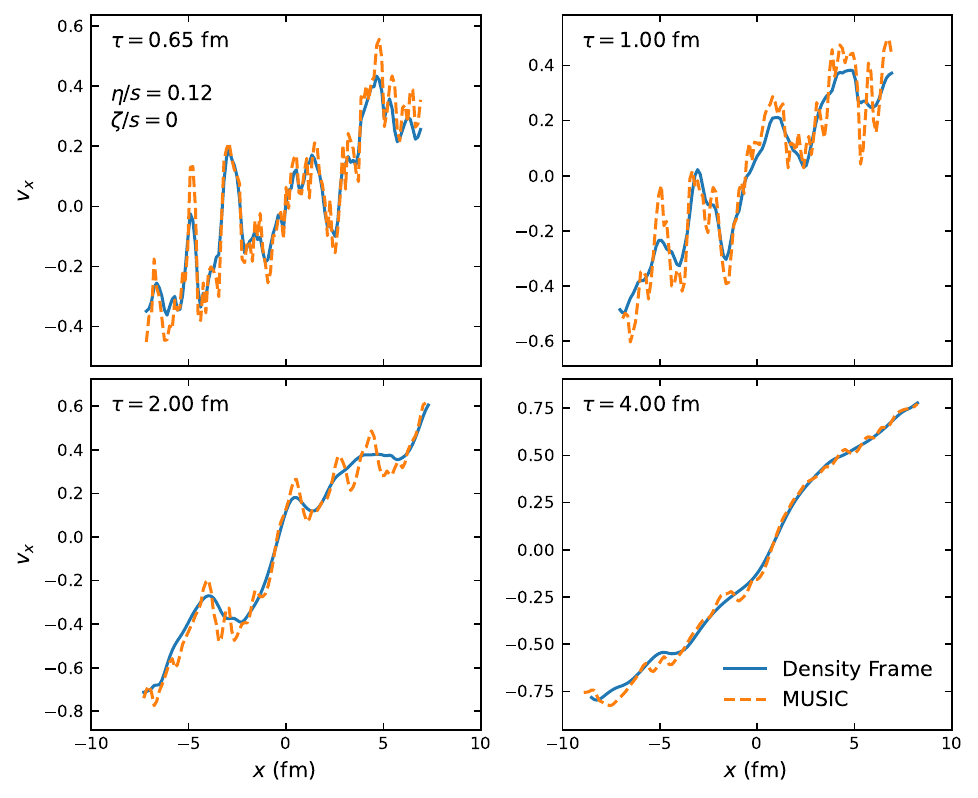}
    \caption{The $x-$component of the transverse velocity along the transverse slice ($y=0$) obtained with Density Frame hydrodynamics (solid blue) and MUSIC (dashed orange) at proper times $\tau=$ 0.65, 1.0, 2.0 and 4.0 fm. Both simulations are initialized with the same IP-Glasma event at $\tau_0 = 0.6$ fm and have $\eta/s=0.12$ and $\zeta/s=0$.}
    \label{fig:vx_1d}
\end{figure}

To facilitate a direct comparison with MUSIC, we transform the Density Frame energy-momentum tensor to the Landau frame at selected times.
Figure \ref{fig:temperature_2d} compares the transverse temperature profiles obtained from Density Frame hydrodynamics and MUSIC at different proper times. 
Although the overall temperature distributions remain similar, the Density Frame formulation dissipates small-scale structures more rapidly, producing noticeably smoother profiles. This difference is particularly evident in the one-dimensional slices at the Cartesian transverse coordinate $y=0$, shown in Fig. \ref{fig:temperature_1d}. A similar smoothing of small-scale fluctuations is observed in the transverse velocity component $v_x$, as illustrated in Figs. \ref{fig:vx_2d} and \ref{fig:vx_1d}.

\begin{figure}[!ht]
    \centering
    \includegraphics[width=0.9\linewidth]{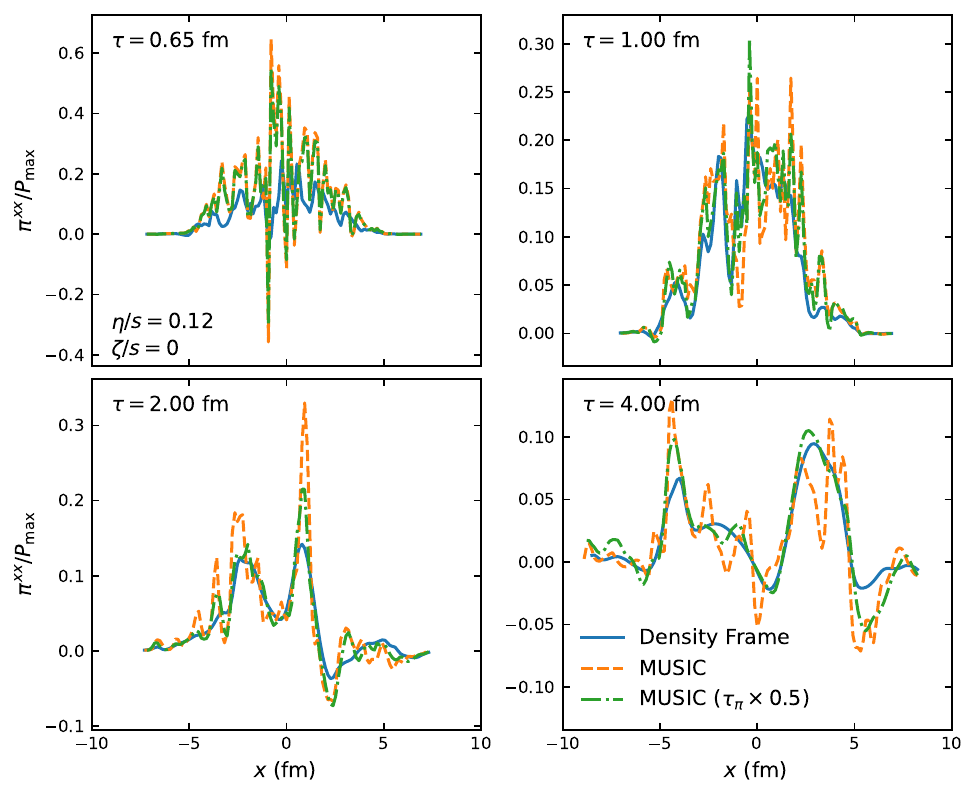}
    \caption{Shear-viscous stress component $\pi^{xx}$, normalized by the maximum pressure $P_{\rm max}$ in each simulation at the corresponding time, along the transverse slice ($y=0$), obtained with Density Frame hydrodynamics (solid blue) and MUSIC (dashed orange) at proper times $\tau=$0.65, 1.0, 2.0 and 4.0 fm. Results from MUSIC with the shear relaxation time reduced by a factor of two, $\tau_\pi\rightarrow\tau_\pi/2$ are also shown (dash-dotted green). Both simulations are initialized with the same IP-Glasma event at $\tau_0 = 0.6$ fm and have $\eta/s=0.12$ and $\zeta/s=0$.}
    \label{fig:pi_1d}
\end{figure}

Figure \ref{fig:pi_1d} compares the shear-viscous stress component $\pi^{xx}$, normalized by the maximum pressure $P_{\mathrm{max}}$ in each setup at each proper time, along the transverse slice $y=0$. We also show MUSIC results with the shear relaxation time $\tau_\pi$ reduced by a factor of two, from its default value of $T \tau_\pi = 5(\eta/s)$ to $T \tau_\pi = (5/2)(\eta/s)$. As with the temperature and velocity profiles, the Density Frame formulation produces a smoother shear-stress profile, with small-scale fluctuations dissipating more rapidly than in MUSIC. Although the hydrodynamic evolution begins at $\tau_0 = 0.6$ fm, differences between the two formulations are already visible at $\tau = 0.65$ fm. In MUSIC, the shear-stress tensor $\pi^{\mu\nu}$ is initialized using values from IP-Glasma and subsequently evolves according to the Israel--Stewart relaxation equations. In contrast, the Density Frame formulation has no independent viscous degrees of freedom, and the shear stress is determined instantaneously by the local hydrodynamic gradients through the constitutive relations. Consequently, the Density Frame shear stress responds immediately to changes in the hydrodynamic fields, whereas the MUSIC shear stress approaches its first-order constitutive value over a finite relaxation time. Reducing $\tau_\pi$ accelerates this relaxation, bringing the MUSIC results closer to those of the Density Frame, which we show in Figure \ref{fig:pi_1d}. 

\begin{figure}[!ht]
    \centering
    \includegraphics[width=0.9\linewidth]{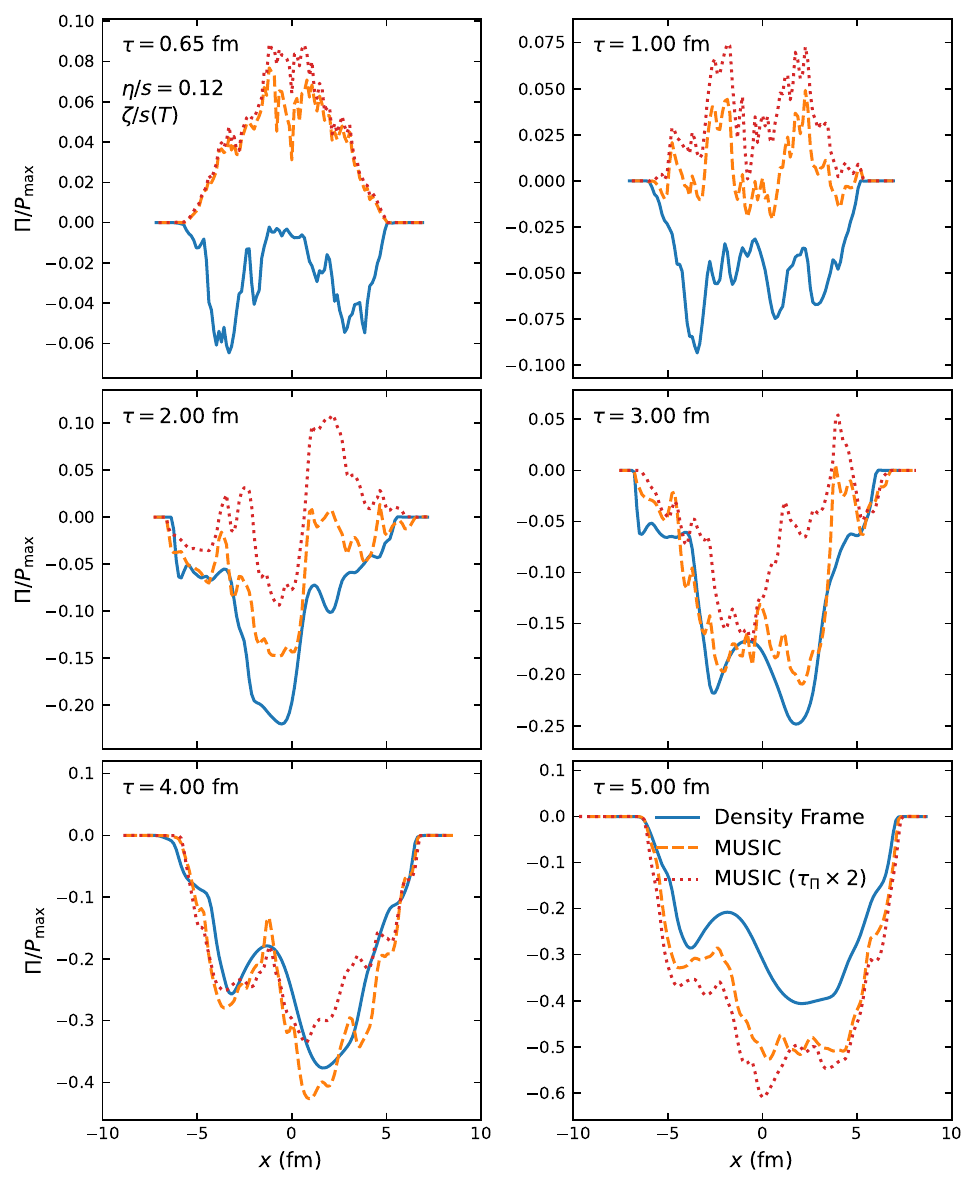}
    \caption{Bulk stress $\Pi$, normalized by the maximum pressure $P_{\rm max}$ in each simulation at the corresponding time, along the transverse slice ($y=0$), obtained with Density Frame hydrodynamics (solid blue) and MUSIC (dashed orange) at proper times $\tau=$0.65, 1.0, 2.0, 3.0, 4.0 and 5.0 fm. Results from MUSIC with twice the bulk relaxation time $\tau_\Pi\rightarrow2\tau_\Pi$ are also shown (dotted red). Both simulations are initialized with the same IP-Glasma event at $\tau_0 = 0.6$ fm and have $\eta/s=0.12$ and a temperature dependent $\zeta/s$.}
    \label{fig:Pi_1d}
\end{figure}

The difference between the two formulations is even more pronounced in the bulk-viscous sector, as shown in Fig. \ref{fig:Pi_1d}. For this comparison, we introduce a temperature-dependent bulk viscosity parametrized in Ref. \cite{Schenke:2020mbo} and show MUSIC results with both the default bulk relaxation time $\tau_\Pi$ given in \cite{Denicol:2014vaa}, and twice its value. As discussed above, in the IP-Glasma + MUSIC framework, the bulk pressure $\Pi$ is initialized to account for the mismatch between the equations of state used in the two stages. 
In the Density Frame, the bulk pressure is determined instantaneously by the local hydrodynamic gradients through the constitutive relations. It therefore takes its first-order hydrodynamic value $\Pi = -\zeta\partial_\mu u^\mu$, which is negative for the expanding system considered here and acts against the expansion, as is generally expected. The MUSIC bulk pressure gradually approaches the Density Frame result over a finite relaxation time, with a larger $\tau_\Pi$ producing a more delayed response.

\begin{figure}[!ht]
    \centering
    \includegraphics[width=1\linewidth]{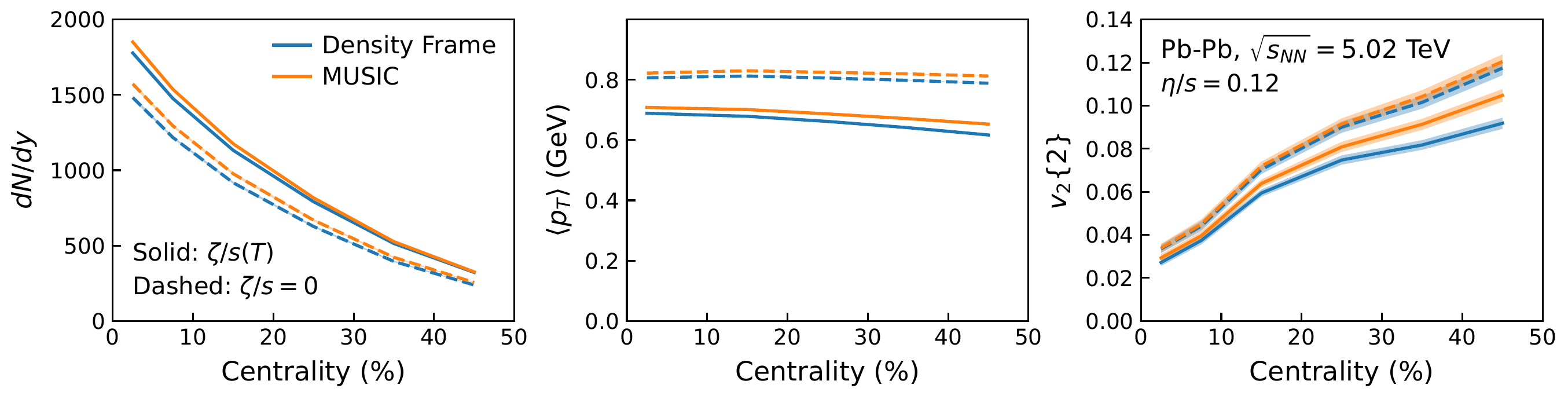}
    \caption{Comparison of charged-particle multiplicity $dN/dy$ (left), mean transverse momentum $\langle p_T\rangle$ (middle) and elliptic flow $v_2\{2\}$ (right) as functions of centrality for Pb-Pb collisions at $\sqrt{s_{NN}}=5.02$ TeV. Results from Density Frame hydrodynamics (blue) and MUSIC (orange) are shown with shear viscosity $\eta/s=0.12$. Solid lines correspond to simulations with temperature-dependent bulk viscosity $\zeta/s(T)$, while dashed lines correspond to zero bulk viscosity.}
    \label{fig:observables}
\end{figure}

\subsection{Observables}
\label{sec:observables}

Finally, we compare the two hydrodynamic formulations at the level of experimental observables
in Pb--Pb collisions at $\sqrt{s_{\mathrm{NN}}}=5.02$~TeV. For this comparison, we simulate 775
IP-Glasma events spanning the 0--50\% centrality range, with the hydrodynamic evolution
initialized at $\tau_0=0.4$~fm. Both formulations use identical model parameters, taken from
Ref.~\cite{Schenke:2020mbo}, rather than parameters calibrated separately to data. We consider both the temperature-dependent bulk viscosity
parametrization employed there and a shear-only scenario in which the bulk viscosity is set to
zero. Figure~\ref{fig:observables} shows the charged-particle multiplicity $dN/dy$, mean
transverse momentum $\langle p_T\rangle$, and elliptic flow coefficient $v_2\{2\}$ as functions
of centrality. Despite the differences in the evolution of the local hydrodynamic fields, the
two formulations yield broadly similar results, particularly in the absence of bulk viscosity.
Including bulk viscosity leads to somewhat larger differences in $v_2\{2\}$, consistent with the
more pronounced discrepancies in the bulk-pressure evolution discussed above.

In more detail, with shear viscosity alone the elliptic flow from the two codes agrees within
the statistical errors, while the multiplicity and mean transverse momentum in the Density Frame
are lower by about 5\% and 2--3\%, respectively. These two shifts are not independent, since the
mean transverse momentum and the multiplicity are correlated. The non-hydrodynamic modes of
MIS therefore have little effect on these observables, consistent with previous Bayesian
analyses, which found only weak constraints on the shear relaxation
time~\cite{JETSCAPE,PhysRevLett.126.202301,Nijs:2020roc}. With bulk viscosity, $v_2\{2\}$ in peripheral
collisions is about 10\% lower in the Density Frame. Even so, the differences between the two
codes are generally smaller than the effect of the bulk viscosity itself (compare the solid and
dashed curves). It is possible that certain choices of specific bulk viscosity could lead to larger differences between MIS and Density Frame. The bulk viscosity  from Ref.~\cite{Schenke:2020mbo},  used in the present calculations, is relatively broad in temperature. If a narrower peak is used (e.g. Ref.~\cite{Ryu:2015vwa}), especially a peak close to the particlization temperature, it leaves less time for MIS to relax to the Navier-Stokes bulk pressure, and difference with Density Frame could be more significant. More studies will be needed to resolve this question.

We emphasize that the two formulations are compared here using identical model parameters,
rather than parameters independently calibrated to experimental data. 
The differences in final state observables between Density Frame and MIS could in principle be reduced further by re-adjusting the model parameters.
The current results suggest that when MIS and Density Frame are both calibrated to experimental data, they can have a similar agreement, although with different constraints on the model parameters.

\section{Conclusions and Outlook}
\label{sec:conclusions}

We have presented a simulation of viscous hydrodynamics in the Density Frame for
heavy-ion collisions in 2+1 dimensions, with a public, documented
code~\cite{density-frame-github} that is suitable for a complete Bayesian analysis
of heavy-ion data. We also provide a modified version of MUSIC that incorporates the Density Frame hydrodynamic solver as an alternative evolution engine~\cite{MUSIC-DensityFrame-github}. We compared the Density Frame results with the exact solution of kinetic theory with Gubser
symmetry (\Sect{sec:kinetic} and \app{app:gubser}), for which we presented a new analysis of how
the kinetic solution matches onto hydrodynamics. We also compared Density Frame with MIS-based MUSIC for
IP-Glasma initial conditions (\Sect{sec:music}), which are known to be numerically challenging. The Density Frame simulations are quite
robust and readily evolve the IP-Glasma initial conditions: the only regulator is a small minimal energy density, imposed at the initial time. The reason
for this stability is that, as in the continuum, the viscous strains follow from variational derivatives of a discrete dissipative function, the positive semi-definite quadratic form that determines the
entropy production (\Sect{sec:entropy}). Consequently, the semi-discrete viscous step produces
entropy on all length scales.

The comparison with MIS illustrates this. Small-scale structure in the initial
conditions is dissipated much more quickly in the Density Frame (see
Figs.~\ref{fig:temperature_2d} and \ref{fig:vx_2d}). In MUSIC the viscous stresses at short
wavelengths are non-hydrodynamic modes, which relax over the time $\tau_\pi$, so
structures smaller than a mean free path survive for a time of order $\tau_\pi$ instead of
being damped on their own viscous time scale. At later times,
however, the two codes give 
very similar
temperature and flow velocity
(Figs.~\ref{fig:temperature_1d} and \ref{fig:vx_1d}), and the shear stresses are largely
compatible once MUSIC has relaxed (Fig.~\ref{fig:pi_1d}). The largest difference is in the
bulk-viscous sector. In the Density Frame the bulk pressure adjusts immediately to the
underlying flow, while in MIS hydrodynamics it relaxes over the time $\tau_\Pi$, which can become long in the
crossover region. As a result, the bulk pressure in MIS lags behind that of the density
frame (see Fig.~\ref{fig:Pi_1d}). Since the initialization of the bulk pressure is rather ad-hoc, the bulk evolution of Density Frame provides a valuable scenario to consider, and can help improve theoretical uncertainties on phenomenological constraints of QCD's bulk viscosity from heavy-ion data.

These differences between Density Frame and MIS have only a modest effect on the final observables
(\Sect{sec:observables}) in our simulations. Without bulk viscosity the two codes agree closely, and with it the
differences are smaller than the effect of the bulk viscosity itself (Fig.~\ref{fig:observables}). 
In general, the difference in observables between MIS and Density Frame will depend on the exact choice of temperature dependence of the specific shear and bulk viscosity, which we will explore in future work.

The Density Frame is causal in the ordinary sense: the response follows the source in time.
It is not causal in the relativistic sense, where the response is confined to the forward light
cone. Relativistic causality generally requires short-distance modes that enforce the required
analytic structure~\cite{Heller:2022ejw,Gavassino:2023mad,Hoult:2023clg}. Our comparisons with kinetic
theory and MIS suggest that such modes do not play a significant role in the evolution, at least
for the observables considered here. Indeed, the Gubser comparison shows that foliation
independence, and hence Lorentz invariance, holds to the expected order in gradients
(\Sect{sec:df}).

The Density Frame is now ready for a complete Bayesian analysis of heavy-ion data. Since it has
no relaxation times or other parameters beyond the viscosities and the equation of state, such
an analysis would constrain the transport coefficients of QCD directly. In the future, the same
framework can be used to search for non-hydrodynamic modes in small systems such as p+Pb and O+O collisions, to
address the theoretical error bars in current Bayesian extractions, and, with noise
added through Metropolis steps~\cite{Basar:2024qxd,Harhoff:2026oxb}, to study stochastic
hydrodynamics in heavy-ion collisions.

\begin{acknowledgments}
We thank J.~Bhambure and D.~Almaalol for their contributions to the initial stages of this work.
D.T.\ is supported by the U.S. Department of Energy, Office of Science, Office of Nuclear Physics,
grant No.~DE-FG-02-88ER40388. M.S.\ and J.-F.P.\ are supported by Vanderbilt University and by
the U.S. Department of Energy, Office of Science under Award Number DE-SC-0024347.
This research used resources of the National Energy Research Scientific Computing Center (NERSC), a Department of Energy User Facility using NERSC award NP-ERCAP 35438.
We acknowledge the use of Claude (Anthropic) and ChatGPT (OpenAI) in preparing this manuscript and in developing
analysis codes.
\end{acknowledgments}

\appendix
\section{Matching the Gubser and Bjorken flows}
\label{app:gubser}\label{app:coords}\label{app:kinetic}\label{app:ic}

This appendix collects known results in one notation. The de~Sitter formulation of Gubser flow
and its ideal and first-order viscous solutions are due to Gubser~\cite{Gubser:2010ze}, with subsequent work by Gubser and Yarom~\cite{Gubser:2010ui}. The exact solution of the RTA Boltzmann equation
with this symmetry is due to Denicol, Heinz, Martinez, Noronha and
Strickland~\cite{Denicol:2014xca,Denicol:2014tha}, and we solve it with the moment method of
Dash and Roy~\cite{Dash:2020zqx}. Second-order hydrodynamic Gubser flow has been studied in
Israel--Stewart theory~\cite{Marrochio2013IS}, in second-order conformal
hydrodynamics~\cite{HattaNoronhaXiao2014a,HattaNoronhaXiao2014b,Behtash2017Gubser}, and
through the slow-roll expansion~\cite{DenicolNoronha2018slowroll}. Our closed-form second-order
result below is a convenient restatement of these.
With the formulas in hand, we find the precise relation between the Gubser solution and the late-time Bjorken attractor through second order in the gradient expansion. 

The flow expands both along the beam and radially in Milne coordinates, but it is at rest in
another set of coordinates. For a conformal system we may rescale the metric,
$\dd\hat s^2 = \dd s^2/\tau^2$. In the coordinates $(\rho,\theta,\phi,\varsigma)$, with
\begin{equation}
  \sinh\rho = -\frac{R^2-\tau^2+r^2}{2\tau R}, \qquad
  \tan\theta = \frac{2Rr}{R^2+\tau^2-r^2}\, ,
\end{equation}
the rescaled metric is de~Sitter space times a line,
$\dd\hat s^2 = -\dd\rho^2 + \cosh^2\rho\,(\dd\theta^2+\sin^2\theta\,\dd\phi^2) + \dd\varsigma^2$.
Our $R$ corresponds to $1/q$ in the notation of Gubser and most of the subsequent
literature~\cite{Gubser:2010ze,Gubser:2010ui,Denicol:2014xca,Denicol:2014tha}.
With our initial conditions, everything depends only on $\rho$. To return to
Milne coordinates we use $T = \hat T/\tau$, $\varepsilon = \hat\varepsilon/\tau^4$ and
$\pi_{\mu\nu} = \tau^{-2}\,\partial_\mu\hat x^\alpha\partial_\nu\hat x^\beta\,\hat\pi_{\alpha\beta}$.
The fluid at rest becomes the flow $u_\tau = -\cosh\kappa$, $u_r = \sinh\kappa$, with
$\tanh\kappa = 2\tau r/(R^2+\tau^2+r^2)$.

The kinetic problem simplifies in the same way. The distribution function depends on $\rho$
and on two conserved combinations of momenta, $f = f(\rho;\hat p_\Omega^2,\hat p_\varsigma)$, and
the relaxation time is $\hat\tau_R = \cRTA/\hat T$. We solve for the moments $L_n(\rho)$ of
Ref.~\cite{Dash:2020zqx}. The first is the energy density, $L_0 = \hat\varepsilon/\Ce$; the
second gives the pressure anisotropy, $L_1/L_0 = (\hat P_L - \hat P_T)/\hat\varepsilon$, and the
shear stress, $\hat\pi^\varsigma{}_\varsigma = \tfrac23 \Ce L_1$. We keep 160 moments.

The initial state is thermal in de~Sitter coordinates at a very early time $\rho_S$,
$f = e^{-\hat p^\rho/\hat T_S}$, and then streams freely. This gives
$\hat\varepsilon = \Ce\Kinf(Q_1R)^3/\cosh^3\rho$, with $Q_1R = \hat T_S\cosh\rho_S$ and
$\hat T_S = 4\Kinf/\pi$. A short calculation 
shows that the distribution function approaches
the Bessel function form in \eq{eq:yield}. In practice we start
at $\tau_S = e^{-4}\min(\tau_H,R)/1000$, and we have checked
that our results are insensitive to the precise
value.  

\label{app:hydro}%
Hydrodynamics simplifies in the same way. The fluid is at rest in de~Sitter coordinates, so
energy conservation becomes one equation for $\hat T(\rho)$~\cite{Marrochio2013IS,Denicol:2014tha},
\begin{equation}
  \frac{\dd\ln\hat T}{\dd\rho} = -\frac23\tanh\rho
  + \frac13\tanh\rho\,\frac{\hat\pi^\varsigma{}_\varsigma}{\hat s\hat T}\, .
  \label{eq:energy_dS}
\end{equation}
The first term is the ideal expansion and the second is the viscous correction. For the stress
we use second-order conformal hydrodynamics~\cite{Baier:2007ix}. Gubser flow has no vorticity,
so only two second-order coefficients appear, $\eta\tau_\pi$ and $\lambda_1$. In a conformal
theory both are proportional to $\eta^2/(e+P)$, and we write
$\eta\tau_\pi = \bar\tau_\pi\,\eta^2/(e+P)$ and $\lambda_1 = \bar\lambda_1\,\eta^2/(e+P)$. For a
classical ideal gas with the RTA collision integral adopted here, $\bar\tau_\pi = 5$ and
$\bar\lambda_1 = 25/7$~\cite{TeaneyYan2013}. On
the static flow, with $\nabla\cdot u = 2\tanh\rho$ and $\sigma^\varsigma{}_\varsigma = -\tfrac43\tanh\rho$,
the stress is
\begin{equation}
  \frac{\hat\pi^\varsigma{}_\varsigma}{\hat s\hat T} =
  \frac43\,\frac{\eta/s}{\hat T}\tanh\rho
  + \frac{(\eta/s)^2}{\hat T^2}\left[-\frac43\,\bar\tau_\pi\,\mathrm{sech}^2\rho
  + \frac89\,(\bar\lambda_1-\bar\tau_\pi)\tanh^2\rho\right].
  \label{eq:pi2}
\end{equation}
The first term is Navier--Stokes. In the bracket, the $\bar\tau_\pi$ term comes from the time
derivative of $\sigma$, and the $\bar\lambda_1-\bar\tau_\pi$ term from the terms quadratic in
$\sigma$.

We solve \eq{eq:energy_dS} as a series in $1/\chi_0$, where
$\chi_0 = R\,T_0/(5\eta/s) = \hat T_0/(5\eta/s)$ and $\hat T_0 = \hat T(\rho=0)$. As in the text,
we write
\begin{equation}
  \hat T = \hat T_0\cosh^{-2/3}\!\rho\,\left[1+\frac{F^{(1)}(\rho)}{\chi_0}+\frac{F^{(2)}(\rho)}{\chi_0^2}+\dots\right],
  \label{eq:That_series}
\end{equation}
and require each correction to vanish at $\rho=0$. Then $\hat T_0$ is the temperature at $\rho=0$
at every order. At first order, $\dd F^{(1)}/\dd\rho = \tfrac{4}{45}\sinh^2\!\rho\cosh^{-4/3}\!\rho$.
Integrating gives Gubser's result~\cite{Gubser:2010ze},
\begin{equation}
  F^{(1)} = -\frac{4g^3}{135}\,{}_2F_1\!\left(\tfrac32,\tfrac76;\tfrac52;-g^2\right),
  \qquad g\equiv-\sinh\rho\, .
  \label{eq:delta1_general}
\end{equation}
The second order is simpler than the first~\cite{HattaNoronhaXiao2014a,HattaNoronhaXiao2014b}.
Feeding $F^{(1)}$ back into the Navier--Stokes term gives $-F^{(1)}F^{(1)\prime}/\chi_0^2$, which
cancels the same term from expanding the logarithm in \eq{eq:energy_dS}. What is left integrates
to powers of $\cosh^2\!\rho = 1+g^2$:
\begin{equation}
  F^{(2)} = \frac{2}{225}\Bigl[(\bar\lambda_1-\bar\tau_\pi)\bigl[(1+g^2)^{2/3}-1\bigr]
  + (2\bar\lambda_1+\bar\tau_\pi)\bigl[(1+g^2)^{-1/3}-1\bigr]\Bigr].
  \label{eq:delta2_general}
\end{equation}
We checked the result against a numerical solution of \eq{eq:energy_dS}.

It remains to fix $\hat T_0$ by matching to the Bjorken regime. At early times at $r=R$,
$g\simeq R/\tau$ is large. There $F^{(1)}$ grows as $g^{2/3}$ up to a constant, $-\aone$, with
$\aone$ given in \Eq{eq:a1a2}, and of $F^{(2)}$ only the $(1+g^2)^{2/3}$ term grows, so
\begin{equation}
  F^{(1)} \simeq b_1\,g^{2/3} - \aone\, , \qquad
  F^{(2)} \simeq b_2\,g^{4/3} - \atwo\, ,
  \label{eq:delta_early}
\end{equation}
with
\begin{equation}
  b_1 = -\frac{2}{15}\, , \qquad b_2 = \frac{2}{225}\,(\bar\lambda_1-\bar\tau_\pi)\, , \qquad
  \atwo = \frac{2\bar\lambda_1}{75}\, .
  \label{eq:b2_general}
\end{equation}
The $\bar\tau_\pi$ pieces of the constant term in \eq{eq:delta2_general} cancel, so $\atwo$
depends only on $\lambda_1$. The early-time solution therefore has the form of viscous Bjorken
flow,
\begin{equation}
  \hat T \simeq \hat T_0\,P\,g^{-2/3}\bigl[1 + b_1 x + b_2 x^2 + \dots\bigr], \qquad
  P = 1 - \frac{\aone}{\chi_0} - \frac{\atwo}{\chi_0^2} + \dots\, ,
  \label{eq:gubser2_early}
\end{equation}
with $x \equiv g^{2/3}/(\chi_0 P)$. Expanding to second order in $1/\chi_0$ at fixed
$g^{2/3}/\chi_0$ reproduces \eq{eq:delta_early}; there are no cross terms. Since $T = \hat T/\tau$,
the prefactor defines the ideal Bjorken temperature at $r=R$,
$T_{\rm bj}(\tau) \equiv \hat T_0 P/(R^{2/3}\tau^{1/3}) = (\Lambda_1^2/\tau)^{1/3}$, and with it
$x = (5\eta/s)/(\tau T_{\rm bj}) = (\tau_H/\tau)^{2/3}$. The bracket is therefore the second-order
Bjorken series \eq{eq:bj_series}, with the coefficients $b_1$ and $b_2$ of \eq{eq:b2_general}. The prefactor $P$
depends only on $\lambda_1$, and by construction $\hat T_0 P = (\Lambda_1R)^{2/3} = R\,T_{\rm bj}(R)$.
Solving for $\hat T_0$ gives
\begin{equation}
  \frac{\hat T_0}{R\,T_{\rm bj}(R)}
  = 1 + \frac{\aone}{\chi_1} + \frac{\atwo}{\chi_1^2} + \mathcal O(\chi_1^{-3})\, .
  \label{eq:T0match2}
\end{equation}
Here we have re-expanded in the opacity $\chi_1$ rather than $\chi_0$. It is notable that,
through second order, the central temperature depends only on $\lambda_1$, while the
early-time Bjorken flow depends only on $\eta\tau_\pi-\lambda_1$~\cite{Baier:2007ix}.

\bibliographystyle{utphys}
\bibliography{refs2d}
\end{document}